\documentclass[a4paper,onecolumn]{emulateapj}

\usepackage{booktabs}

\usepackage{latexsym}
\usepackage{enumerate}
\usepackage{amstext}
\usepackage{apjfonts}
\usepackage{amsmath}
\usepackage{amssymb}

\usepackage{natbib}
\usepackage{rotating}
\usepackage[colorlinks=true,linkcolor=blue,citecolor=blue]{hyperref}
\usepackage{changepage}
\usepackage{scrextend}
\usepackage[ampersand]{easylist}
\usepackage{xcolor}
\usepackage{ulem}
\usepackage{listings}
\usepackage{soul}
\usepackage{xparse}

\NewDocumentCommand{\scnum}{ >{\SplitArgument{1}{e}}m }
 {\scnumaux#1}
\NewDocumentCommand{\scnumaux}{ m m }
 {#1\,\mathrm e\,{#2}}

\begin{document}

 \title{Cinematic Visualization of Multiresolution Data: Ytini for Adaptive Mesh Refinement in Houdini}

\author{ Kalina Borkiewicz\altaffilmark{1}, J.~P.~Naiman\altaffilmark{1,2}, Haoming Lai\altaffilmark{1,3}}
\altaffiltext{1}{Advanced Visualization Laboratory, National Center for Supercomputing Applications, 1205 West Clark Street, Urbana, IL 61801}
\altaffiltext{2}{Harvard-Smithsonian Center for Astrophysics, The Institute for Theory and Computation, 60 Garden Street, Cambridge, MA 02138}
\altaffiltext{3}{Brown University, HCI Group, 115 Waterman Street, Providence, RI 02912}

 \begin{abstract} 
 
We have entered the era of large multidimensional datasets represented by increasingly complex data structures. Current tools for scientific visualization are not optimized to efficiently and intuitively create cinematic production quality, time-evolving  representations of numerical data for broad impact science communication via film, media, or journalism. To present such data in a cinematic environment, it is advantageous to develop methods that integrate these complex data structures into industry standard visual effects software packages, which provide a myriad of control features otherwise unavailable in traditional scientific visualization software. In this paper, we present the general methodology for the import and visualization of nested multiresolution datasets into commercially available visual effects software. We further provide a specific example of importing Adaptive Mesh Refinement data into the software \textit{Houdini}. This paper builds on our previous work, which describes a method for using \textit{Houdini} to visualize uniform Cartesian datasets. We summarize a tutorial available on the website \textit{www.ytini.com}, which includes sample data downloads, Python code, and various other resources to simplify the process of importing and rendering multiresolution data.

\end{abstract}


\section{Introduction}

Data visualization is defined as the display of information in a graphical format, and can refer to different types of data (e.g. relational and spatial) and different styles of graphics (e.g. two-dimensional graphs and three-dimensional renderings). Either type of visualization serves two purposes: data analysis and communication. Visualization and traditional numerical analysis are complementary methods of analyzing relationships between variables, or spotting regions of interest in large datasets \citep{goodman2012}. Additionally, visualization is a way to communicate findings - whether in academic papers, presentations to peers, or when communicating complex concepts to the general public \citep{vogt2016, barnes2008, punzo2015, borkiewicz2017}.

In this paper, we focus on cinematic data visualization for purposes of science communication. The aim of cinematic visualization is not only to be educational and compelling, but also aesthetically pleasing and entertaining in order to have broader appeal. \cite{pandey} suggest that data visualization is more persuasive than communicating with tables or numbers, and \cite{cawthon} have found that aesthetic visualizations are more educational than unattractive ones. By leveraging the familiar visual language established by Hollywood films, a cinematic presentation of science thus creates  interest in topics that may otherwise be thought of as dull and difficult to learn \citep{arroio2010,serra2008,dubcek2003}. As eluded to in \cite{chen}, aesthetic is a critical and as yet unsolved problem in data visualization, growing only more important as data becomes more complex \citep{important1,important2}.

Providing scientists with the means to generate cinematic imagery from their data allows them to to create more impactful, educational, pleasing, and broad-reaching imagery by taking a more aesthetic approach to visualization. Free online platforms such as YouTube or Vimeo can be used directly by scientists to share their visualizations with the broader public. \cite{welbourne} show that user-generated science content (i.e. videos created by scientists or science enthusiasts) is far more popular on YouTube than videos made by professional science communicators. While a cinematic approach to visualization is not yet part of the day-to-day workflow of a typical scientist, it can be used at the end of a research project to communicate with the general public through teaching or outreach, to communicate with peers through presentations and publications, and to communicate with funding agencies. Cinematic data visualization is also of use to visual effects designers who desire to incorporate real science into film  \citep[e.g.][]{interstellar, visart}. This paper focuses on methodology specifically applicable to astronomers. Though cinematic visualization is computationally expensive, looking to the future, this cost will decrease as technology and techniques improve.

\subsection{Current State of Visualization Tools} \label{section:currentState}

Several established and well-adopted tools already exist for data visualization (\textit{ParaView}\footnote{https://www.paraview.org/} from \cite{paraview}, \textit{VisIt}\footnote{https://visit.llnl.gov/} from \cite{visit}, \textit{yt}\footnote{http://yt-project.org/} from \cite{turk2011}) and likewise for visual effects and animation (\textit{3DS Max}\footnote{https://www.autodesk.com/products/3ds-max/}, \textit{Blender}\footnote{https://www.blender.org/}, \textit{Houdini}\footnote{https://www.sidefx.com/}, \textit{Maya}\footnote{https://www.autodesk.com/products/maya/}). However, neither category of tools is well-suited for cinematic data visualization. Table \ref{table:software} compares a selection of software at-a-glance, and a discussion in greater detail is provided in the subsections below.

\subsubsection{General-Purpose Scientific Tools}\label{section:scitools}

\textit{ParaView}, \textit{VisIt}, \textit{yt}, and other scientific tools are widely used by the community for data visualization. The primary purpose of these tools is to support data analysis capabilities for scientists, and visualization is only a feature that enhances these capabilities. In addition to their prevalence and ability to analyze and visualize data, these tools have many features that are desirable to scientist users, including the ability to natively read multiple complex scientific data types, parallelizability for running on clusters or supercomputers, and extensibility through open-source codes. These tools create images that are familiar and understandable to the experts who frequently view them, however they can be unintuitive to non-scientists, or scientists from different fields. Because visualization is a secondary concern to analysis, these software do not put heavy emphasis on employing best practices for visual communication. One example of this is the prevalence of rainbow color maps in these software tools. Research has shown that rainbow color maps are perceptually confusing \citep{rainbowmap1, rainbowmap2}, but nevertheless many science tools continue to include them in their default pre-set color options, and scientists continue to use them even if they are aware of their problems, out of convenience and convention \citep{rainbowmap3}.

While it is possible to create visually compelling imagery using scientific tools, features that are critical for cinematic aesthetics are often hidden behind menus in favor of promoting data analysis features, or simply missing from the interface altogether. Many complex lighting, shading, coloring, and camera controls or algorithms do not exist in these software, and application of these advanced computer graphics techniques must be done either externally or programmatically. The list of the most prominent non-scientific visual effects features that remained underdeveloped in the majority of scientific software includes:
\begin{easylist}[itemize]
& Color: \textit{yt} allows for the speed of choosing between predefined colormaps, but there is no interface for extending to complex transfer functions. \textit{VisIt} and \textit{ParaView} allow for editing colormaps, but this feature is hidden in favor of selecting from a list of predetermined options.

& Lighting: Most scientific tools provide limited lighting options. For example, \textit{VisIt} allows the user to select between only three light types (as compared to \textit{Houdini}'s twelve): using the camera as a light source, using the data object as a light source, or using ambient lighting. The color and brightness of the light can be changed, but not the intensity, radius of influence, or shadow. The direction of the light can be set by typing in a vector with values between -1 and 1, rather than by being able to click and drag the focal point to the desired location.

& Camera: Many tools do provide camera movement options, however the extent of this is limited. For example, users can easily create cameras that move or zoom linearly in \textit{ParaView}, but there is no interface for creating more complex camera choreography, which may ease in and out, follow an interesting portion of the data as it evolves, and move organically in a non-linear way, as if it were recorded by a live cinematographer.

& Contextualization: In scientific software, the three-dimensional environment in which the user interacts is constructed around each specific dataset, making the contextualization of the data or combination of multiple datasets difficult. The spatial integration solutions provided in industry standard special effects software help to tackle this, one of the most important unsolved problems in scientific visualization \citep{laramee2014}.

& Cinematography: a computer animation that mimics recording with a physical camera - with effects such as field of view, depth of field, and motion blur - can feel more realistic and cinematic. These features are often lacking from scientific software as they add no scientific value, but they are important in cinematography.
\end{easylist}

\subsubsection{Bespoke Multiresolution Tools}\label{section:bespoketools}

Several customized visualization solutions exist for astronomical multi-resolution data \citep{bespoke1, bespoke2, cpuamr}. While this category of tools and algorithms puts the most emphasis on rapidly and efficiently rendering large multiresolution datasets, they have many of the same basic limitations as scientific tools, as described above - they do not allow for cinematic camera choreography, lighting, or shading, and do not allow for combining multiple datasets. Furthermore, these solutions are not accessible to a wide variety of users as they are specialized, and unlike the other categories of tools, tend to not be supported by large user communities or customer support teams.

\subsubsection{Visual Effects Tools}\label{section:vfxtools}

\begin{figure*}
\centering
\includegraphics[width=1.0\textwidth]{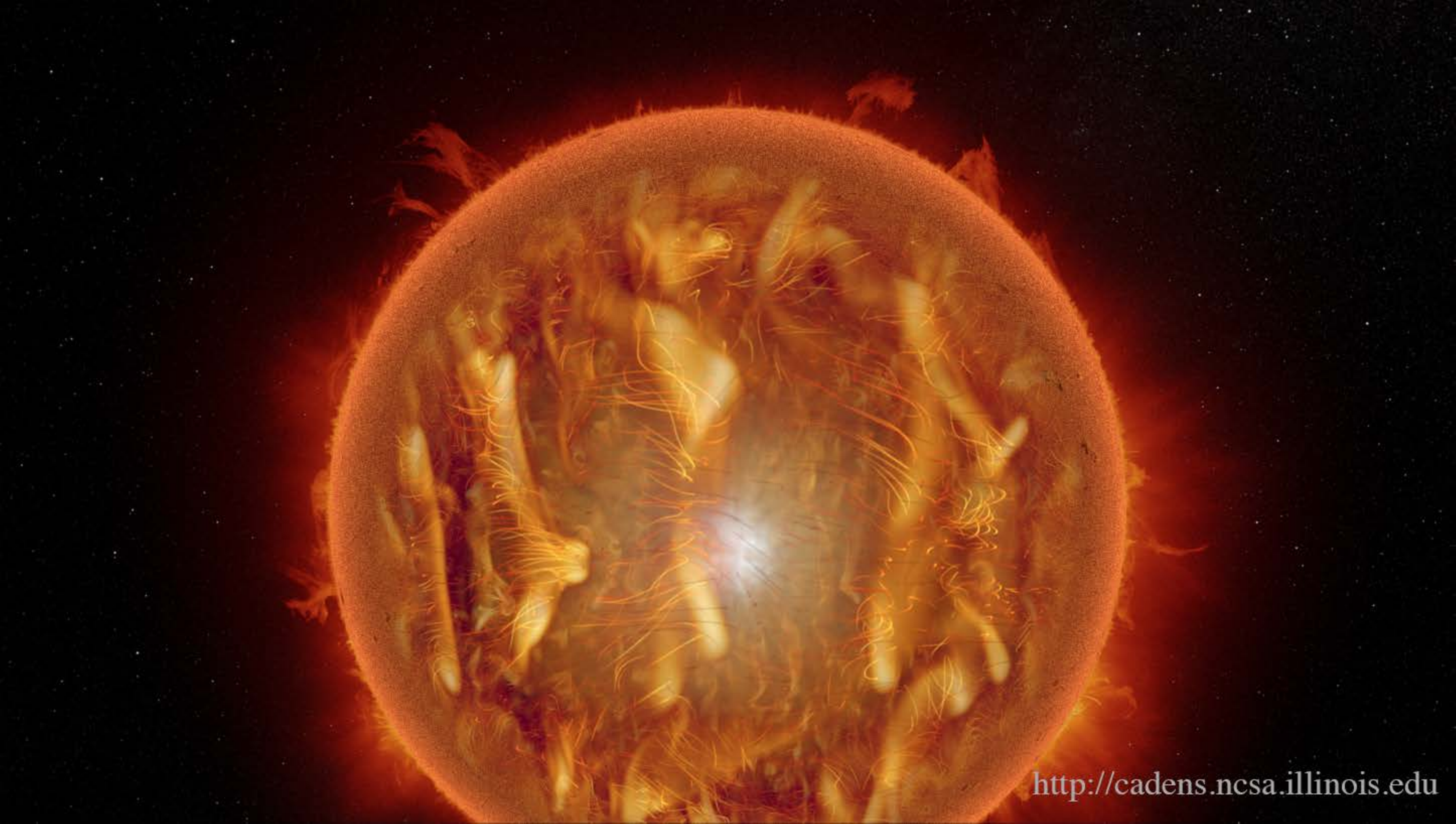}
\caption{ Rendering created in \textit{Houdini} combining multiple datasets - a volumetric solar interior simulation \citep{miesch}, a solar surface simulation \citep{rempel}, solar rim imagery \citep{sdo}, and background star imagery \citep{allsky} - along with a statistical core and streamlines traced through the solar interior volume, using the \textit{Houdini} software. This image is not rendered with the multiresolution workflow described in this paper but is intended as an example of features in \textit{Houdini}.}
\label{fig:sun}
\end{figure*}

Visual effects software excels in many use cases described above where scientific and bespoke visualization tools fall short. Visual effects tools work under the paradigm of designing a ``scene'' rather than visualizing a dataset. A single scene can consist of multiple datasets of any type - volumes, surface geometry, particles - alongside designed artistic elements, for easy contextualization. For example, Figure \ref{fig:sun} shows a visualization of the Sun created in \textit{Houdini} which combines a solar surface simulation, a volumetric solar interior simulation, streamlines traced through the volumetric data (traced using \textit{Houdini}), an artistic volumetric core, photographic imagery along the sun's perimeter, and a background of stars.

In addition to self-illuminating data as described in the example above, any number, position, and type of realistic external lighting can be added. For example, \textit{Houdini} provides easy click-and-drag access to the following twelve lighting types: point light, spot light, area light, geometry light, volume light, distant light, environment light, sky light, indirect light, caustic light, portal light, and ambient light.

Furthermore, visual effects tools provide a plethora of camera controls. Camera paths can be created either by keyframing (selecting a start and end position in time with automatic smooth interpolation), clicking-and-dragging, or tying the camera's view to a specific object so that the camera follows the object as it moves or evolves. There is a variety of camera path interpolation algorithms to choose between, and two-dimensional graphs of the camera positions and rotations can be manually adjusted. Camera motions can ``ease in'' and ``ease out'' to give the viewer the feeling that a physical camera is being operated by a cameraperson, rather than algorithmic. Many camera effects such as lens flare, motion blur, depth of field, field of view, film grain can easily be added to further make the imagery cinematic.

It is also worth noting here that, as discussed in section \ref{section:scitools}, in contrast to scientific packages, color and opacity maps are highly customizable (refer to Figure \ref{fig:colormap}).

However, visual effects software is not designed for use by scientists and their complex datasets, but is instead designed to meet the needs of artists who only need their data to be visually plausible and can suffice with simple, uniform, Cartesian data representations. These software can natively read spreadsheets and other simple file formats, but custom data readers, plugins, translators, or data converters must be written in order to import more complex data types.

\subsubsection{Scientific Visualization Plugins into Visual Effects Software}\label{section:plugins}

Recent work at the intersection of science and cinema aims to partially address this divide between categories of tools by creating scientific plugins into the 3D computer graphics software, \textit{Blender} \citep{kent2015,taylor2015,naiman2016}. \textit{Blender} is more accessible to researchers than some other artistic tools for approaching cinematic visualization due to being free and open-source. However, it is not currently the typical tool used by visual effects professionals in cinema, and thus the reverse is not true - plugins such as these do not reach artists to help them approach scientific visualization.

Our previous work with \textit{Ytini}\footnote{http://www.ytini.com/} \citep{naiman2017} integrates the data reading capabilities of the scientific analysis and visualization Python package, \textit{yt}, with the production-quality visual effects software package, \textit{Houdini}.  \textit{Ytini} leverages \textit{Houdini}'s Python interface to connect with the scientific analysis and visualization Python package, \textit{yt}. \textit{yt} is used as a data reader, as it is capable of natively reading a number of simulation and observational data formats, and provides a method for loading generic NumPy arrays to support non-standard formats. Once the data is loaded with \textit{yt}, another Python package, PyOpenVDB, is used to convert the data into the \textit{Houdini}-readable OpenVDB format. \textit{Ytini} generates a uniform grid which can be read and rendered by \textit{Houdini} within a traditional volume rendering workflow. This paper builds on that methodology, introducing a method of importing and rendering nested, multiresolution data into high-end visual effects software.

\subsection{Traditional Volume Rendering} \label{section:volumeRendering}

To render a dataset into an image or a series of images, the data must be imported into the visual effects software's three dimensional space and light rays cast from light sources into a virtual camera. As this process forms the foundation of our updated methods, we describe it here in some detail.

In computer graphics, domain-filling spatial data is often represented in what is called a ``volume''. When volumes are brought into a visual effect software's three dimensional space, they are embedded in the larger domain of the ``scene'', allowing for multiple volumes to be integrated. A standard volume is a three-dimensional grid that is divided into uniform cubes called ``voxels'', which correspond to what are called ``grid cells'' in the scientific community.  Because all voxels within a volume must be the same size, the resolution of the entire volume must be high enough such that the areas of interest in the data are shown with enough detail. Storing data in a traditional volume is thus memory inefficient in the case where the volume has regions that are sparsely filled.

OpenVDB \citep{museth2013} is a memory-efficient sparse volumetric data format, where areas of voxels with a value of zero take up minimal memory. Memory is saved by using a tree structure in which leaf nodes are collapsed into their parent and all nodes in a branch contain the same value (e.g. zero). The OpenVDB tree topology consists of four levels with decreasing branching factors restricted to powers of two.\footnote{Refer to section 2 and Figure 3 of \cite{museth2013} for more details on the OpenVDB data structure.} The method described in this paper uses OpenVDB volumes for memory efficiency and speed, though traditional volumes could be used as well. OpenVDB offers a smaller memory footprint and faster data processing for sparse volumes, and offers benefits like improved CPU cache performance even for dense grids.\footnote{http://www.openvdb.org/documentation/doxygen/faq.html\#sCompareVDB}

\begin{figure*}
\centering
\includegraphics[width=1.0\textwidth]{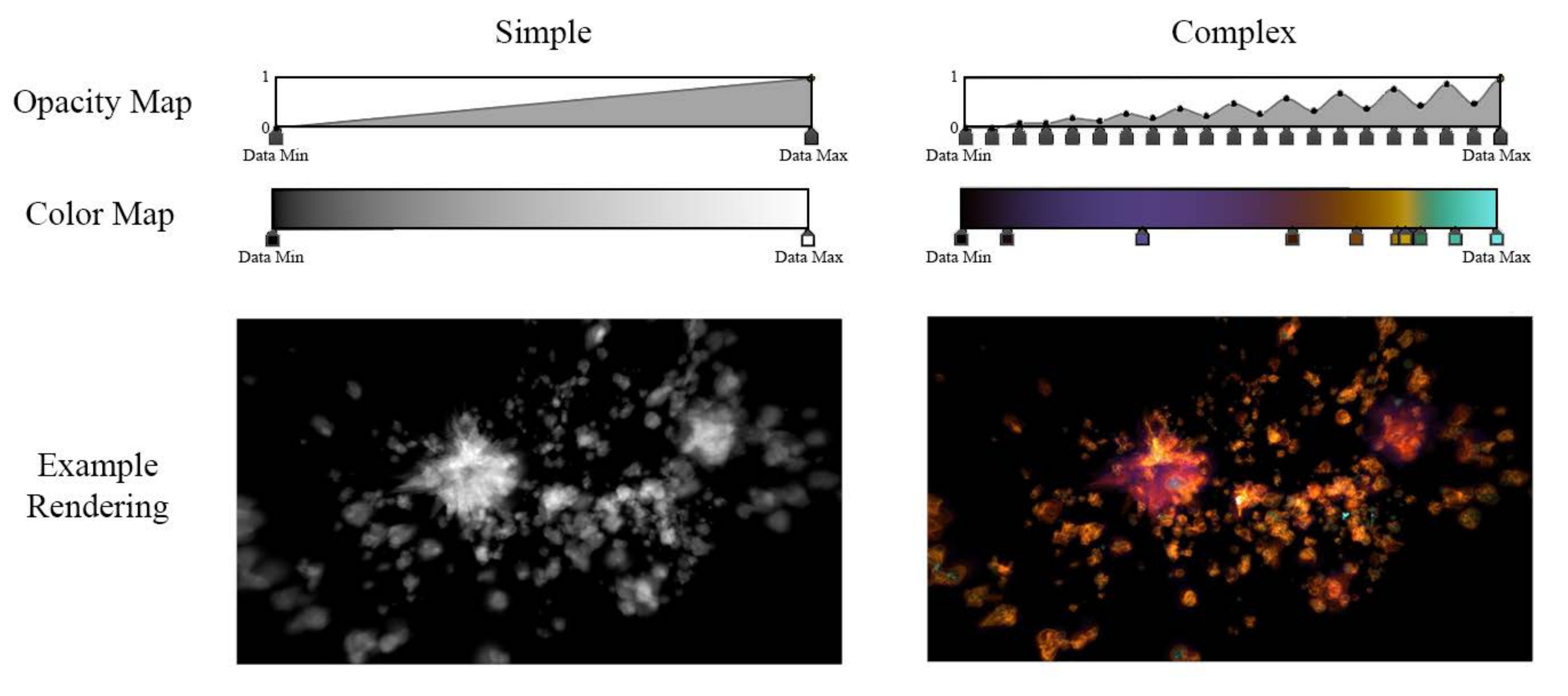}
\caption{ Two different sets of opacity and color maps are applied to the metallicity field in a dataset from \cite{oshea2015}. Left: A simple opacity map changes linearly from 0 to 1 across all data values (top left). Similarly, a simple color map changes from black (RGB 0,0,0) to white (1,1,1) (center left). These are the default opacity and color maps provided by \textit{Houdini}. Right: A more complex opacity map (top right) and color map (center right) can change between many values and colors, and use various interpolation schemes. As with Figure \ref{fig:sun}, this imagery is an example of a feature in \textit{Houdini} and is not rendered with the multiresolution workflow described described in this work.}
\label{fig:colormap}
\end{figure*}

\begin{figure*}
\centering
\includegraphics[width=1.0\textwidth]{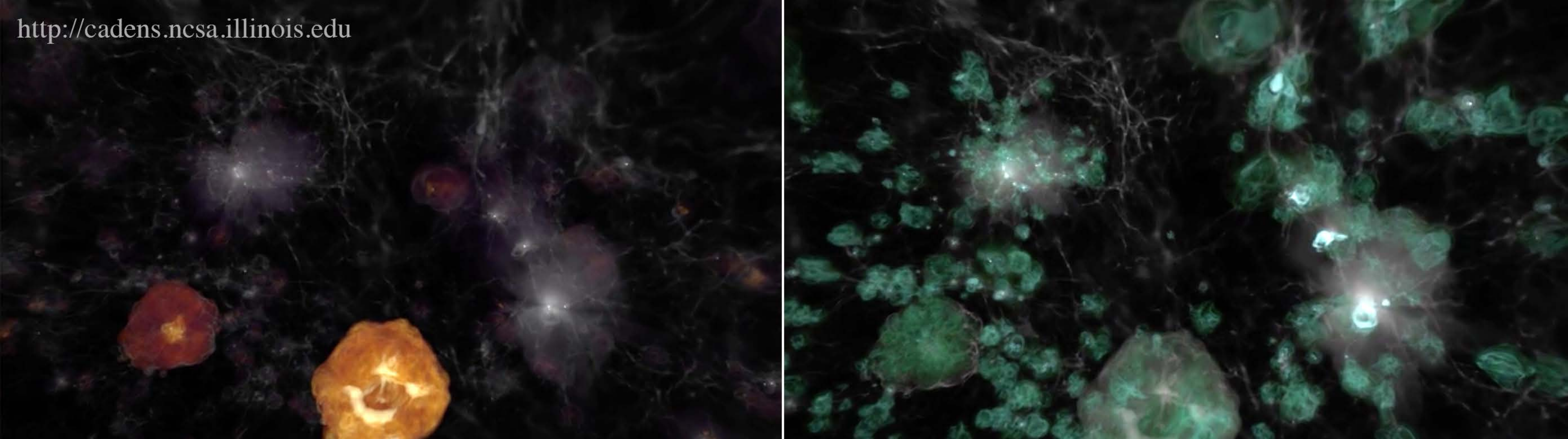}
\caption{ An emissive volumetric dataset from \cite{oshea2015}, where intensity and wavelength of the emitted light is adjusted to highlight different aspects of the data. Both images show the same six variables from a similar camera position --  HI density, HII density, temperature, photo gamma, metallicity, and stars. The left image highlights regions of high temperature in orange while dimming the other attributes and removing their color to maintain context. The right image highlights areas of high metallicity and dims the rest. As with Figure \ref{fig:sun}, this imagery is an example of a feature in \textit{Houdini} and is not rendered with the multiresolution workflow described in this work.}
\label{fig:renvars}
\end{figure*}

To turn a grid of data values into something visually useful, one must assign physical attributes to the data values. A ``material shader'' is used to map data ranges to opacity and color as well as light emission, which is necessary in the case of self-luminescent astrophysical data. In the simplest case, the lowest data value is mapped to transparent and black (opacity and color channels all set to 0), the highest data value is mapped to opaque and white (opacity and color channels all set to 1), and all intermediate values are mapped linearly between the respective 0 and 1 values, to create a grayscale image. An advanced color map assigns many colors and levels of transparency to different data ranges, and use various forms of interpolation, to bring out features in the data. Examples of color maps as created in the software \textit{Houdini} applied to data from \cite{oshea2015} are shown in Figure \ref{fig:colormap}. As evidenced by comparing the left and right panels of Figure \ref{fig:colormap}, the grayscale map emphasizes the high data values while the complex map is able to accentuate multiple interesting data ranges. Astrophysical data often does not require an external light source because gas, stars, galaxies, supernovae, and other such objects have emissive properties. In the case of emissive objects, the intensity and wavelength of the emitted light can be adjusted by the scene designer. For example,  Figure \ref{fig:renvars} shows two snapshots of a visualization that cycles through several volumetric variables of a dataset. The left image highlights regions of high temperature in orange while dimming the other attributes and removing their color to maintain context. The right image highlights areas of high metallicity and dims the rest. In scenes where data is not self-illuminating, for example where a planet requires light from its star in order to be visible to a viewer, a separate light source object must be added into the visual effect software's three dimensional space. Various types of external lights can be added to the scene to highlight features and/or cast shadows, as described in section \ref{section:vfxtools}. One scene can host a multitude of light sources - for example, stars represented as point lights can be embedded within emissive gas.

\begin{figure*}
\centering
\includegraphics[width=1.0\textwidth]{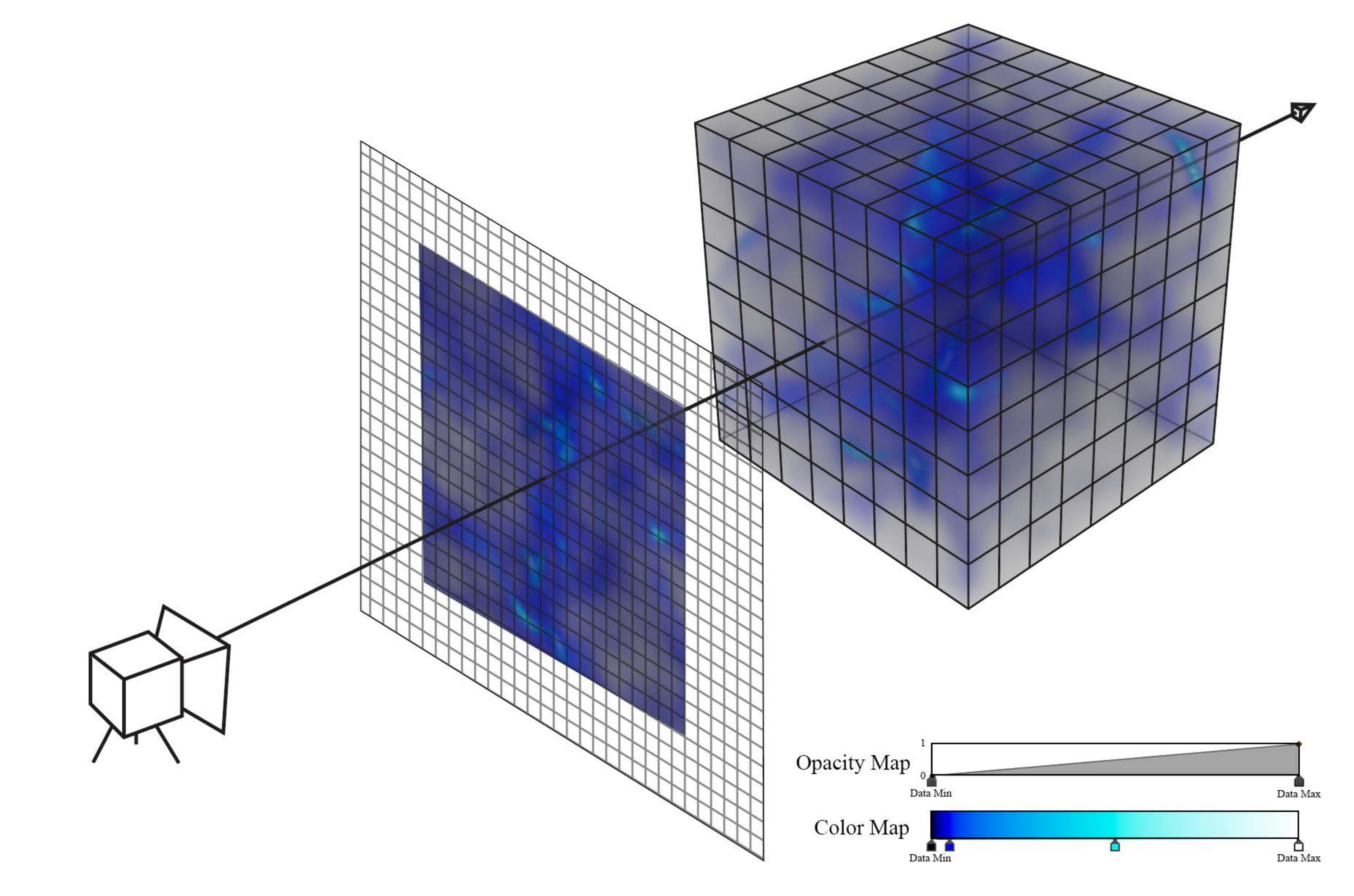}
\caption{ Description of a raytracing process, shown with the dataset from \cite{enzo2013} that is used in the online tutorial. Each ray starts at a single camera position, passes through a 2D plane of pixels which will be the resulting image, and traverses through a translucent volume. Points are sampled through the volume along the ray. Each sample point has certain characteristics, such as color, opacity, and illumination, which are described by the shader. The result of the combination of these sample point characteristics is placed at the pixel on the 2D plane through which the ray passes. }
\label{fig:raytracing}
\end{figure*}

Once data objects are imported and the ways in which lighting sources interact with these data structures are defined, the process of generating an image from the space and data can begin. The process of converting a three-dimensional scene and the volumes and elements therein into a two-dimensional image is called "rendering". "Raytracing", in the nomenclature of the visual effects community, is a common rendering technique, which reverses the process of human vision. \footnote{In this context, when we reference raytracing, we are referring to the rendering technique within the visual effects software, and not the process by which radiation sources can add energy and momentum into a hydrodynamical simulation \citep[e.g.][]{raytracingastro}. While it is possible to replace or extend \textit{Houdini}'s default renderer, Mantra, with a custom raytracer, that is beyond the scope of this paper. As an example, this has been done to render the black hole in the film ``Interstellar'' \citep{interstellar}. } In a simplified description of human vision, light originates at a light source, bounces off an object, and scatters in many directions; some of these scattered rays reach our eyes, allowing us to see the object. This is an inefficient process for a computer to mimic. Raytracing reverses this process, and a simulated ray of light originates at the eye (or camera), is traced through a 2D image plane of pixels, and traverses through the volume in the three-dimensional scene. The renderer takes samples at points along the ray, and uses the shader to map the resulting color and opacity to the pixel on the image plane through which the ray crosses, as depicted in Figure \ref{fig:raytracing}.  

\begin{figure*}
\centering
\includegraphics[width=1.0\textwidth]{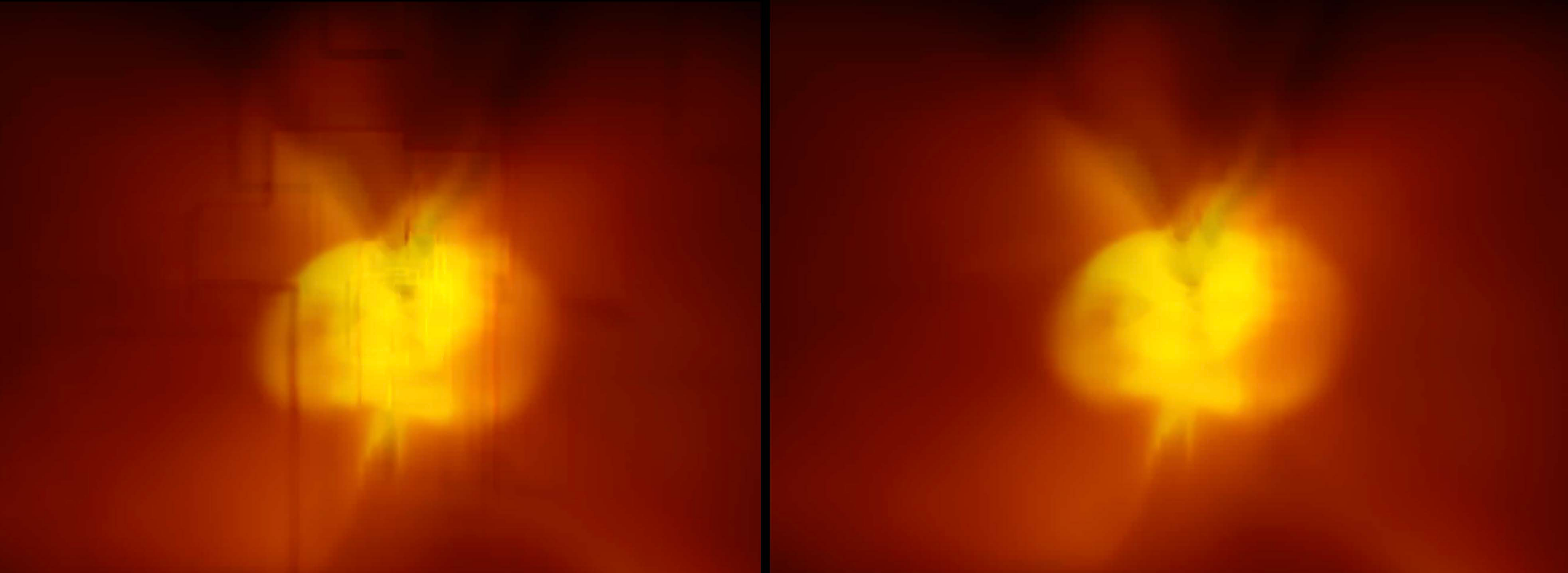}
\caption{Two renderings of the density field of the dataset from \cite{wise}. The image on the left shows edge effects resulting from placing volumes of different voxel sizes next to one another without further adjustment. The image on the right shows our method of data preprocessing and shading. }
\label{fig:blog}
\end{figure*}

The use of multiresolution data complicates this picture.  Multiresolution data necessitates volumes with different sized voxels to be placed next to each other or in nested sets. However, the customary treatment of volumetric data in traditional visual effects software produces a multitude of edge effects which create artifacts within the rendering of the data, as shown in the left panel of Figure \ref{fig:blog}.  To address this, we have developed new methods of data pre-processing and a modification to the default volumetric interpolation methods used during the raytracing process within industry standard special effects software.

\section{Challenges with Multiresolution Volumes in Visual Effects Software} \label{section:amrVolumeRendering}

Irrespective of the particular visual effects software package used, the visualization of multiresolution data is met with several general challenges - including how the data is imported from disk, how each data point is represented spatially within the software, how nested regions are treated, and how the edges between regions of differing resolutions are rendered.  Here, we outline these issues conceptually. We provide an in-depth description of the methods for processing and importing multiresolution data into \textit{Houdini} in section \ref{section:amrHoudini}.

Visual effects software typically requires all volumetric data to be in uniform, rectangular  grids.  Thus, the first step in processing multiresolution data is to subdivide the data into segments across which all voxels are uniformly sized. The simplest implementation of this separates the higher-resolution areas from the lower-resolution grids in which they are contained, and creates uniform volumes out of each resulting fragment. However, this may potentially create hundreds of volumes, which are difficult to manage when each volume needs to be assigned a shader in the software. The preferable method is to use a sparse volumetric data format such as OpenVDB, which is natively supported in \textit{Houdini}, available in \textit{Maya} via a plugin, and can be installed manually for other software packages. The use of a sparse volume allows one to combine areas with the same voxel size into a single volume, regardless of the separation between them.

An important consideration when reading data into volumes is how the data is positioned and whether the source multiresolution data is vertex- or cell-centered, and whether this is how the software expects data to be represented. \textit{Houdini}, for instance, uses a cell-centered method of representing volumes, but a vertex-oriented method for positioning the volume within its three-dimensional space, incorrectly placing the vertex of a volume at its (0,0,0) origin. If these data models and software expectations are not in alignment, the data may have to be shifted so that it will be interpreted correctly by the software.  This can either be done in the processing step from raw data to OpenVDB voxel data before import, or within the visual effects software itself.

Once data storage and representation have been considered, the volumetric data may be loaded into the visual effects software. If data is present at the same position in space in multiple higher- and lower-resolution volumes, one must select only the highest available resolution data to display. Overlapping regions create unwanted areas of higher density and brightness in the final rendered image due to one data value being accounted for multiple times. During the shading and rendering processes, each volume is treated as a separate entity, without the ability to receive information about the data in surrounding volumes or objects. Thus to remove overlapping regions, each volume must consist of not only the data values, but also an associated mask derived from any data nested within the grid, such that there is a clearly-defined boundary in which to place the nested volume within the domain of the current volume.

Finally, one must account for the visual effects software's treatment of edges between volumetric data of differing resolutions. In the typical artistic volume rendering use case, the user desires to hide the underlying grid structure of the data to create smooth volumes where the discrete voxels appear to be continuous. Thus, the software automatically interpolates data values between cells. However, with multiple nested volumes of differing resolutions, this leads to unwanted edge artifacts where the volume fades into the background at the edges.  In the case where masks are applied to the data as discussed in the previous paragraph, further care needs to be taken such that interpolation of both the data and the mask do not lead to dark borders at the data and mask edges.

\section{Multiresolution Volumes in Houdini: Adaptive Mesh Refinement Data} \label{section:amrHoudini}

To demonstrate our novel process of importing multiresolution data as volumes into visual effects software, we use examples of Adaptive Mesh Refinement (AMR) data in \textit{Houdini}. AMR \citep[e.g.][]{amr} is a method of subdividing a domain within a hydrodynamical simulation such that regions of interest are of higher resolution (as determined a priori by the scientist or with some criteria such as regions with large gradients in the fluid's density and pressure). Data is broken up into ``levels'' of refinement; the level with the lowest refinement is referred to as \textit{level 0} and each subsequent level has a resolution that increases by a specified factor, often 2. Each level is made up of one or multiple ``grids'', regions where the data is clustered. The AMR data formats used by Enzo \citep{enzo2013}, Athena \citep{athena2008}, and FLASH \citep{flash2000} are explicitly supported in the Python code provided in the Appendix, though this methodology can be applied to other formats and more generic multiresolution nested data as well. 

Here we outline the developed technique for reading and rendering nested, multiresolution AMR data in \textit{Houdini} and highlight some problems that can arise from an improper treatment in each step of the process. We provide pseudocode in section \ref{section:pseudocode}, and leave a discussion of the practical step-by-step implementation to section \ref{section:workflow}. Aside from the first and last steps, the order in which these steps are performed is arbitrary; the sequence as described here was selected in order to group conceptually related steps together.

\begin{figure*}
\centering
\includegraphics[width=0.9\textwidth]{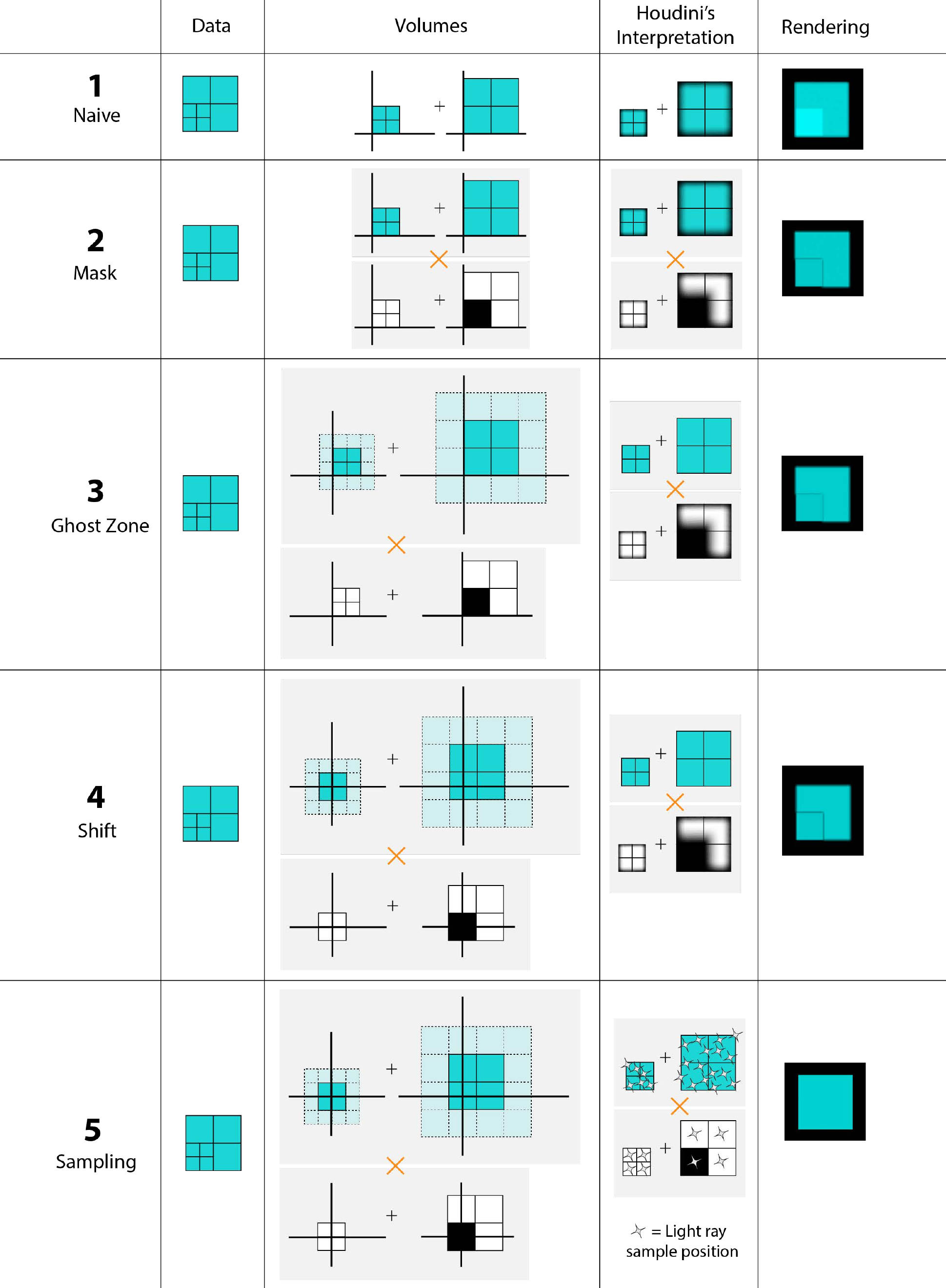}
\caption{ Visual description of the steps needed to visualize AMR data in \textit{Houdini}. The first step in our method converts the grid levels into OpenVDB volumes and places them in \textit{Houdini}'s three dimensional environment. This is a naive implementation due to the visible areas of higher density and brightness where the volumes overlap. Step 2 extracts a mask of the overlapping regions, and multiplies each data value by the mask value. This removes the overlap but causes gaps between the levels and shows that the data is misaligned. Step 3 adds ghost zone voxels around the data to lessen the gap between levels, making sure to keep the original data value positions the same. Step 4 aligns the data at the edges by shifting each level by an additional 1/2 level-0-sized voxel. The final step involves creating a shader in \textit{Houdini} to sample the mask value by index rather than by position in order to prevent automatic edge interpolation, thus removing the remaining edge artifacts. Standard sampling occurs at all steps, but is only shown in this figure in step 5 where it is altered.} See text for further details.
\label{fig:diagram}
\end{figure*}

The first step in this process requires representing the AMR data in terms of uniform volumes. \textit{Houdini} natively supports the sparse OpenVDB volumetric data format, allowing for the usage of memory-efficient volumes. Thus, our method begins by generating a separate volume out of each AMR level, throughout which the cell size (and thus the voxel size) is constant. A single AMR level may be made up of many spatially separated grids, with a constant cell size across all grids in a level. \textit{yt} is used to retrieve the AMR data level-by-level in lines 29-30 of the Python code provided in the Appendix. All grids are subsequently combined into a single uniform volume in lines 35-36. The use of the sparse storage of each level of data in OpenVDB volumes efficiently stores the active data regions, and minimal information about the unpopulated regions. A naive import of each of these OpenVDB volumes results in a double counting of rendered volumes as shown in diagram 1 of Figure \ref{fig:diagram}. This issue arises from the fact that both the low and high refinement levels are rendered in the image at each point that they overlap.

To avoid such overlapping of data, volumetric masks are produced to inform the shader of regions where there is both low and high resolution data. A mask associated with each level has a value of 1 at a voxel where there is no higher resolution data, and 0 where another further level of refinement is present. The mask is obtained from the data in line 32 of the Python code. Although an additional mask value is exported for every data value, the volume file size does not double. The mask volumes are compressed with OpenVDB and are extremely lightweight, often only requiring tens of kilobytes.  A shader is created in \textit{Houdini} to multiply the data value at each voxel by the mask value, resulting in the selection of the highest available level of refinement at any point sampled by the renderer, as the low resolution data is masked out. As shown in Figure \ref{fig:diagram}, diagram 2, while this solves the problem of overlapping data, it creates a new issue - borders between levels, revealing the AMR structure.

Several factors contribute to the creation of these borders. The first issue involves the renderer's interpolation of data values between cells. The purpose of this interpolation is to create volumes that look smooth, and do not show the underlying grid structure. This works well for the standard use case of rendering a single, standalone volume. However in this scenario, this results in the outermost voxels interpolating with zero and creating unwanted edge artifacts. To interpolate data values up to the boundaries of the edge cells, the data grids are extended one cell beyond the edge of the grid, and an extra ``ghost zone'' voxel is appended to the volume. A \textit{yt} function is utilized to retrieve these ghost zone voxels, which uses the AMR structure to interpolate the data values from the adjacent cells in the lower or higher resolution grids. This occurs in line 31 of the Python code. The ghost zone voxels cause the values at the edge of the volume to interpolate to the ghost zone values rather than to zero. The mask associated with each level as described in the paragraph above does not extend to cover these extra voxels to ensure that the ghost zone voxels are not rendered.  This effectively causes the data to be multiplied by 0 where the mask does not exist. The mask is not explicitly padded with ghost zone values of 0, as this would introduce border artifacts where grids within the same level are directly next to one another. The effect of adding ghost zones to the data is shown in Figure \ref{fig:diagram}, diagram 3.

Circumventing such edge interpolation addresses one of the issues causing borders at refinement level boundaries, however the borders will still persist until the data levels are properly aligned in the three dimensional space.  The data is shifted in the pre-processing of data, in lines 44-49 of the Python code. \textit{Houdini} positions each volume such that the vertex of the first cell is at the origin (0,0,0). To account for the additional ghost zone voxel around the perimeter, the grids are first translated by subtracting one level 0-sized voxel from \textit{Houdini}'s placement position of each level's volume. This places the vertex of the first ghost zone cell of each volume at (\textit{-level0VoxelSize, -level0VoxelSize, -level0VoxelSize}), and the vertex of the first data cell at (0,0,0). Next, the data is shifted further to place the center of the cell at the origin, as opposed to its vertex. This requires translating the data and the mask by 1/2 voxel for each respective level's voxel size. Each level's data volume is translated by -1/2 voxel and its mask volume by +1/2 voxel to maintain a difference of 1 voxel between the two volumes, to account for the additional ghost zone voxels added to the data volume but not the mask volume. The final required amount of shift for the data along the x, y, and z dimensions is \textit{-voxelSize/2 - level0VoxelSize}, and the shift for the mask is \textit{voxelSize/2 - level0VoxelSize}. Figure \ref{fig:diagram}, diagram 4 shows that this shift actually appears to increase the visibility of the border between the levels, however, notice that the left and bottom edges of the rendering are now aligned.

Finally, the remaining persistent border is due to the treatment of the mask edge effects by \textit{Houdini} and must be addressed within the shader itself. Applying the mask to the data in the shader as opposed to removing data from the volumes directly allows for design flexibility in determining how many AMR levels to show from a specific camera position, as showing a smaller number of levels may not be visually discernible from a distant camera, thus saving time and memory on data loading and rendering. Once the volume is imported into \textit{Houdini}, a customized shader is created to interpret the AMR data. With a standard volume shader, the raytracer sends a ray that samples a position and returns interpolated color and opacity values at that position. To avoid \textit{Houdini}'s interpolation across cell boundaries for the mask, each sample position is first used to look up the corresponding index in the mask grid. The value of the mask at that index is retrieved, rather than the value at the ray's exact sampling position, to ensure that the renderer uses the uninterpolated center of the cell containing a mask value of exactly 0 or 1. The data value is multiplied by this mask to determine which of the overlapping volumes to sample at that position. Thus only one volume will have a mask value of 1 at any given sampled position, which creates a seamless rendering, as shown in Figure \ref{fig:diagram}, diagram 5.

\subsection{Pseudocode}\label{section:pseudocode}

The implementation of the algorithm as described above can be summarized using the following pseudocodes. The first section describes the data processing step shown in rows 1-4 of Figure \ref{fig:diagram}. The Python implementation of this pseudocode is provided in the Appendix and at \textit{ytini.com}.
\\
\\
\lstset{} 
\begin{lstlisting}
amrData = yt.readData("/path/to/mydata")

for every amrLevel in amrData {
	vdbData = vdb.initiateGrid()
	vdbMask = vdb.initiateGrid()

	for every amrGrid in amrLevel {
		amrMask               = yt.getMask( amrGrid )
		amrGridWithGhostZones = yt.getGhostZones( amrGrid )
		amrGridStartIndex     = yt.getStartIndex( amrGrid )

		vdb.copy( from=amrGridWithGhostZones, to=vdbData, xyz=amrGridStartIndex ) 
		vdb.copy( from=amrMask,               to=vdbMask, xyz=amrGridStartIndex )
	}

	currentVoxelSize = amrData.getCellSize( level = amrLevel )
	largestVoxelSize = amrData.getCellSize( level = 0 )

	vdbData.setVoxelSize( currentVoxelSize )
	vdbMask.setVoxelSize( currentVoxelSize )

	vdbData.translateXYZ( -currentVoxelSize/2 - largestVoxelSize )
	vdbMask.translateXYZ(  currentVoxelSize/2 - largestVoxelSize )

	vdb.writeVolume( [vdbData, vdbMask], "/path/to/output_level#.vdb" )
}
\end{lstlisting}

The pseudocode below describes the custom shader, which determines what level .vdb is selected for rendering at every position in the virtual scene, shown in row 6 of Figure \ref{fig:diagram}. This is implemented in \textit{Houdini}'s node-based visual programming language. 
\lstset{} 
\begin{lstlisting}
for every sampleAlongRayTracedLightRay {
	if sampleAlongRayTracedLightRay.intersects( amrVolume ) {
		position = sampleAlongRayTracedLightRay.getPosition()
		volumeColorAtPosition   = standardRaytracerFunctionToGetColor()
		volumeOpacityAtPosition = standardRaytracerFunctionToGetOpacity()
	
		index =  amrVolume.getCellIndexAtPosition( position )
		maskValue = amrVolume.getMaskValueAtIndex( index )
		volumeOpacityAtPosition = volumeOpacity * maskValue

		return volumeColorAtPosition, volumeOpacityAtPosition
	}
}
\end{lstlisting}

\subsection{Practical Workflow}\label{section:workflow}

For users interested in leveraging the provided code rather than developing a custom implementation of the algorithm described above, we present the practical workflow for importing and rendering a sample AMR dataset into Houdini. For a detailed step-by-step tutorial, refer to the Appendix, or \textit{ytini.com}.

\begin{enumerate}

\item{Run the Python script provided in the Appendix and at \textit{ytini.com}, modifying it to point to your own dataset. This will create a .vdb file for each level in your AMR dataset.}

\item{Download the AMR shader from \textit{ytini.com} and import in into \textit{Houdini}.}

\item{Import the .vdb files into Houdini, each level into its own geometry.}

\item{Import the AMR shader into \textit{Houdini}. Create one shader for each level, and assign each shader to its respective geometry. Only one shader needs to be maintained, the others should be linked as reference copies.}

\item{Modify the main shader parameters such as color, opacity, emission, etc. as desired, and render the image.}
\end{enumerate}

\section{Benchmarking Results}\label{section:results}

The memory efficiency of the outlined methodology is demonstrated in comparisons against the alternative implementation of AMR in visual effects software, where data is instead represented as a single uniform grid. Numerical comparisons are provided in Tables \ref{table:comparisonSmall} and \ref{table:comparisonLarge} for two datasets - the small sample data from \cite{enzo2013} used in the online tutorial and made up of three AMR levels, as well as a larger dataset from \cite{wise} with thirteen AMR levels. All tests are done by rendering the density fields of each respective dataset. The uniform grids were created using yt's ``Covering Grid'' function\footnote{https://yt-project.org/doc/examining/low\_level\_inspection.html\#examining-grid-data-in-a-fixed-resolution-array}, which creates a uniform gridding of the domain at a specified resolution. It was not possible to represent the \cite{wise} dataset in the form of a single uniform grid at a resolution higher than the fourth refinement level, due to memory constraints. Partitioning the data could have made representation at a higher level possible, but this would have only resulted in a longer render time and higher memory footprint in terms of file size and voxel count, thus comparing the rendering of thirteen multiresolution levels using our method against a uniform representation at four levels is assumed to be sufficient to show the benefits of our method. In Table \ref{table:comparisonLarge}, all thirteen multiresolution levels are represented in the ``Ytini Method'' column, against four levels of the ``Uniform Grid'' column. The smaller dataset has a full representation of the data in both columns of Table \ref{table:comparisonSmall}.

The memory footprint is reduced for both the small and large datasets. The file size and voxel count of the small dataset are reduced by factors of $\sim$6 and $\sim$10 respectively, and $\sim$34 and $\sim$24 for the large dataset despite the limited comparison using different levels of refinement (four levels against thirteen levels). Voxel count would be reduced by a factor of $\sim$$\scnum{3.25e9}$ for the large dataset if a uniform representation at the highest refinement level were possible, and we expect the other values would scale similarly. Data loading time is increased by a factor of $\sim$3 for the small dataset, and decreased by a factor of $\sim$33 for the large dataset. The render memory and render time remains approximately consistent for the smaller dataset, and is reduced by factors of $\sim$6 and $\sim$4 for the larger dataset for a camera outside of the domain. Again we emphasize the comparison isn't complete given that we are comparing four against thirteen levels. For a camera inside the domain, viewing the most highly-defined region, render time is similar between the fully-resolved thirteen-level AMR \textit{Ytini} representation as compared to the four-level uniform grid representation, and render memory is reduced by a factor of $\sim$6.

For the small dataset, a few results presented in Table \ref{table:comparisonSmall} are similar between the two data representation methods, or the \textit{Ytini} method gives worse performance. The reasons for the slight increase in render time (factor of $\sim$1.1) and render memory (factor of $\sim$1.01) are due to the fact that an additional mask lookup step must occur at each raytracing sample position. The data loading time is increased by a factor of $\sim$3 due to the reading of two volumes - the data and the mask - as compared to just the data in the uniform grid representation, though both methods produce volumes that load in a matter of milliseconds. However, in even a slightly larger dataset, the relative effect of these additional steps becomes small, and the \textit{Ytini} method quickly begins to outperform the uniform grid method.

\begin{figure*}
\centering
\includegraphics[width=1.0\textwidth]{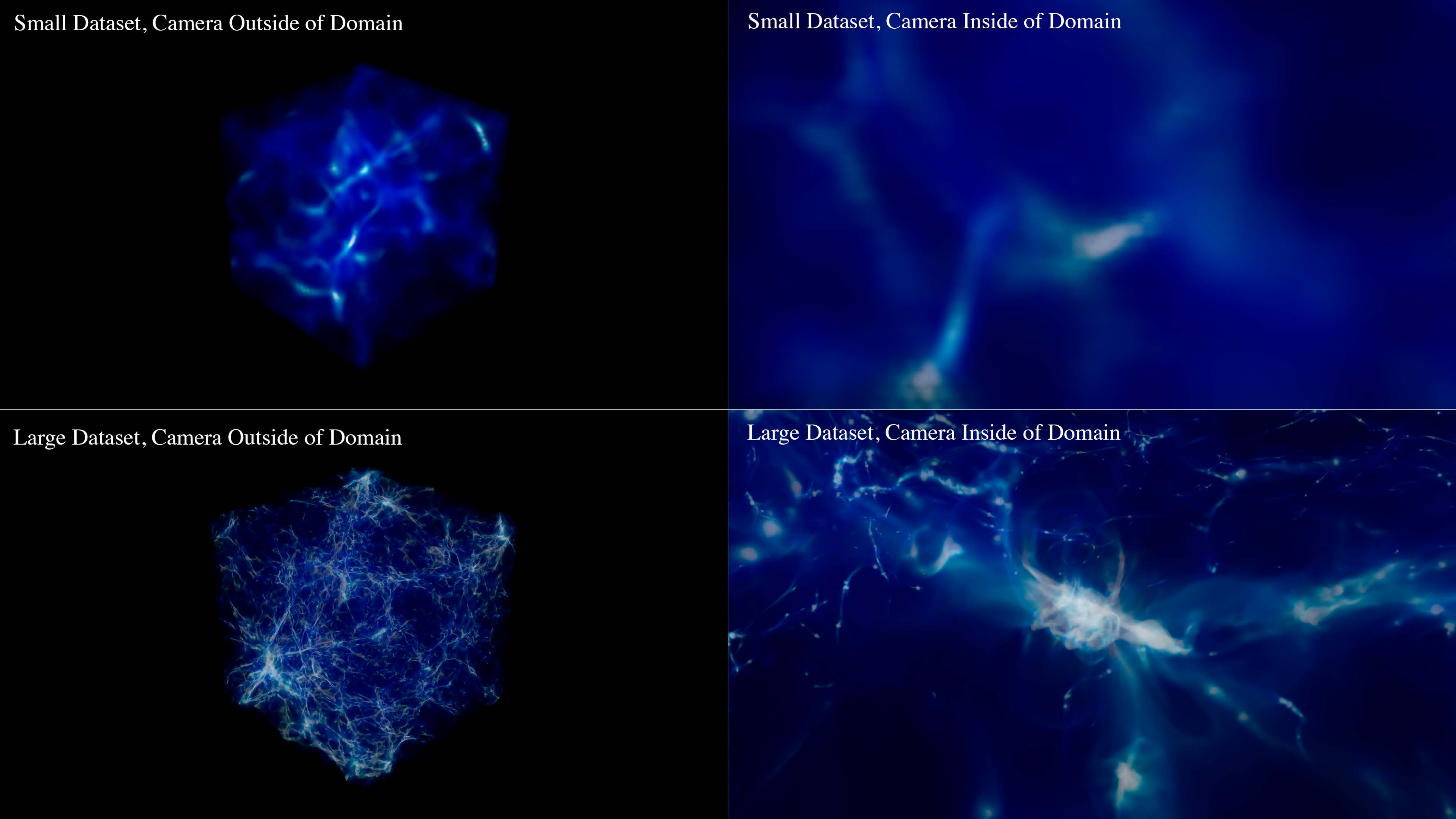}
\caption{Top row: Rendered images that produced the results in Table \ref{table:comparisonSmall}, showing the density field of the dataset from \cite{enzo2013}. Bottom row: Rendered images that produced the results in Table \ref{table:comparisonLarge}, showing the density field of the dataset from \cite{wise}. Both images on the right are viewing the most highly-resolved region of each dataset.}
\label{fig:renderings}
\end{figure*}

Specific memory storage, read times and render times will depend on the specific dataset and its AMR structure, design decisions, as well as the computer used to render the image. The results in Tables \ref{table:comparisonSmall} and \ref{table:comparisonLarge} are an average of three runs performed on a Dell T7600 workstation, with 80GB of RAM and one 8-core 2.4GHz E5-2665 Sandy Bridge processor. The images in this test were rendered at a resolution of 1920x1080 with a camera placed outside the domain of the data, as well as inside the domain of the data, observing a highly-defined region, as shown in Figure \ref{fig:renderings}. Design decisions such as image resolution, lighting, and camera placement can affect render time, but not data reading time or data memory requirements. For a uniform grid data representation, render time scales linearly with the number of pixels rendered and number of samples along each raytraced light ray; with multiresolution data, a higher resolution image is additionally likely to require accessing higher refinement levels, and the change in render time depends on the exact viewpoint. When a camera is placed inside the domain, render times increase for both the uniform grid representation as well as our multiresolution representation. In the edge case where the camera is placed inside the domain at a position such that a feature resolved at the highest level fills the entire frame, the render time of our method approaches the render time of the uniform grid representation at that same level. Though render time increases, memory savings are still retained; thus our method is still preferable to the alternative.

The Python code provided in the Appendix is a simple implementation that maps the domain of the data to one \textit{Houdini} ``unit''. It is possible for a dataset which maps to a small voxel size and contains a high number of levels to result in errors in the code relating to floating point precision. In this case, a simple multiplier can be applied to the voxel size to scale the domain so as to be contained in more than one \textit{Houdini} unit. We have been able to represent as many as 25  AMR levels using this method, before the code results in errors once again. The \cite{wise} dataset in Table \ref{table:comparisonLarge} was scaled up by a factor of 2\textsuperscript{4} in this manner.

For large hydrodynamical simulations, memory often becomes the limiting factor in rendering images due to the physical memory limitations of computer hardware. If render time is a concern, there are several ways to reduce it within the \textit{Houdini} interface, such as lowering the default parameter values for ``Pixel Samples'', ``Volume Quality'', and/or ``Stochastic Samples'' in the \textit{Mantra} node in the \lstinline[basicstyle=\ttfamily]|/out/| network. Additionally, both render time and memory can be improved by thresholding data in the Python preprocessing step to discard data with little to no impact on the final render (e.g. extremely low density data). The workflow and methodology described here provide a memory-efficient means to import complex datasets into industry standard special effects software.

\section{Summary and Future Plans} \label{section:summary}

Advancements in the field of computer graphics, as driven by Hollywood films, raise the public's expectations for what constitutes quality imagery across all domains - from films, to gaming, to science. Giving the scientific community the ability to make use of visual effects tools will allow for the creation of higher-fidelity visualizations that meet the high bar set by modern cinema. The need for cinematic visualization is two-fold: 1) Experts can make use of these tools to better communicate with those not in their field, and 2) Scientific visualization will look increasingly anachronistic if we cannot keep pace with advancements in the arts. In this work we introduced a method for importing and rendering nested multiresolution datasets in visual effects software.  The technical steps to format and import multiresolution data are as follows, with a detailed summary provided in section \ref{section:amrHoudini} with Figure \ref{fig:diagram}:
\begin{enumerate}
\item{Separate each level of refinement and represent it as a uniform sparse volume.}
\item{Export a mask volume for each level to indicate to the renderer where higher-resolution grids contain data at any given position.}
\item{Write an additional ghost zone voxel around each data grid to prevent dark borders between the nested grid levels due to automatic falloff to zero at the grid edges.}
\item{Shift the data and mask volumes so that the data is properly aligned with the origin in the software's three dimensional environment.}
\item{Finally, ensure that the mask values do not automatically interpolate but remain constant throughout a voxel by sampling the mask volumes by index rather than by position in the material shader.}
\end{enumerate}

\indent As a concrete example of this technical prescription, we describe a workflow for importing, shading, and rendering a sample AMR dataset within the user interface of the software \textit{Houdini} in the Appendix. Sample Python code is provided for reading the data and exporting OpenVDB volumes. A detailed tutorial with downloads is available at \textit{www.ytini.com/tutorials/tutorial\_amr.html}. 

Our ongoing work focuses on extending support to other multiresolution data formats and AMR codes, including particle support. We also plan to further optimize render time by omitting highly defined regions if they do not contribute to the rendered image, based on camera position. Progress will be routinely updated on the website. 
\\
\\

The authors would like to thank Stuart Levy, AJ Christensen, Donna Cox, Bob Patterson, Jeff Carpenter, Mark Van Moer, Nathan Goldbaum, and Matt Turk of NCSA; Mark Elendt and Jeff Lait of SideFX; Sebastian Frith; Thomas Robitaille; and the paper's anonymous referee. This work is supported by NSF award for CADENS ACI-1445176 and NSF grant AST-1402480. 

\clearpage

\begin{table}
\centering
\caption {Comparison of Visual Effects vs Scientific Software}
\label{table:software}
\begin{tabular}{lccc|ccccccc}
\hline \hline
 & Houdini & Maya & Blender & yt & ParaView & VisIt & Vapor & IDL & Glue & AstroPy\footnote{AstroPy does not have built-in visualization functionality; two-dimensional visualization can be done using the matplotlib library.} \\
\hline
Data Analysis                            &   &   &   & X & X & X & X & X & X & X \\
Modular Coding Paradigm\footnote{\textit{Houdini} differs from most other visual effects software due to its modular, procedural, code-like GUI paradigm, making it more data-friendly. See Section 2 of our first \textit{Ytini} paper \citep{naiman2017} for more information.}           
                                         & X &   &   & X & X & X & X & X & X & X \\
Supports Multiple Scientific Data Formats&   &   &   & X & X & X &   & X & X & X \\
Open Source                              &   &   & X & X & X & X & X &   & X & X \\
Graphical User Interface                 & X & X & X &   & X & X & X &   & X &   \\
Renders Volumes                          & X & X & X & X & X & X & X & X & X &   \\
Renders Surfaces                         & X & X & X &   & X & X & X & X &   &   \\
Renders Multiple Geometry Types in Scene & X & X & X &   & X & X & X &   & X &   \\
Renders Multiple Datasets in Scene       & X & X & X &   & X & X & X &   & X &   \\
Visual Transfer Function Editor          & X & X & X &   & X & X & X & X &   &   \\
Complex Lighting\footnote{Lighting options beyond emissivity or a small number of rudimentary lighting types (such as point lighting), with few adjustable settings beyond color or intensity.}
                                         & X & X & X &   &   &   &   &   &   &   \\
Complex Shading, Material Effects        & X & X & X &   &   &   &   &   &   &   \\
Interactive Camera Controls\footnote{A camera refers to an object with a viewpoint that is separate from the user's view.}                & X & X & X &   &   &   &   &   &   &   \\
Interactive Design of Elements/Effects   & X & X & X &   &   &   &   &   &   &   \\
\end{tabular}
\\
Comparison of several visual effects tools (left) and scientific visualization tools (right). The benchmark is based on whether the software natively and intuitively supports these features, though workarounds may be possible.
\\
\end{table}

\begin{table}
\centering
\caption {Comparison of Methods with a Small Dataset}
\label{table:comparisonSmall}
\begin{tabular}{lcc}
\hline \hline
Metric & Uniform Grid & Ytini Method \\
\hline
Number of Voxels & 2,097,152 & 213,785 \\
File Size & 4.19~MB & 0.67~MB \\
Data Loading Time & 70.53~ms & 205.19~ms \\
\hline
Render Memory (outside camera) & 1.53~GB & 1.55~GB \\
Render Time (outside camera) & 9~s & 10~s \\
\hline
Render Memory (inside camera) & 1.41~GB & 1.38~GB \\
Render Time (inside camera) & 46~s & 50~s \\
\end{tabular}
\\
Comparison of the memory efficiency and speed of our method against a uniform grid data representation using the density field of the three-level dataset from \cite{enzo2013}, with two camera positions - outside the domain of the data, and inside the domain, observing one of the highest-defined regions. The camera positions of the renderings are shown in the top half of Figure \ref{fig:renderings}.
\end{table}

\begin{table}
\centering
\caption {Comparison of Methods with a ``Large'' Dataset}
\label{table:comparisonLarge}
\begin{tabular}{lcc}
\hline \hline
Metric & Uniform Grid (4 Levels) & Ytini Method (13 Levels) \\
\hline
Number of Voxels for Levels 0-3 * & $\scnum{8.59e9}$ & $\scnum{2.89e8}$ \\
Number of Voxels for Levels 0-12 * & $\scnum{1.15e18}$ & $\scnum{3.55e8}$ \\
\hline
File Size & 28.59~GB & 0.84~GB\\
Data Loading Time & 4~m : 22~s & 0~m : 8~s\\
\hline
Render Memory (outside camera) & 37.72~GB & 6.69~GB\\
Render Time (outside camera) & 17~m~:~15~s & 4~m~:~34~s\\
\hline
Render Memory (inside camera) & 37.72~GB & 6.52~GB\\
Render Time (inside camera) & 28~m~:~03~s & 27~m~:~41~s\\
\end{tabular}
\\
Comparison of the memory efficiency and speed of our method against a uniform grid data representation using the thirteen-level dataset from \cite{wise}, with two camera positions - outside the domain of the data, and inside the domain, observing the highest-defined region. The first two rows, marked with an asterisk (*), are a direct comparison of the two methods. For all other rows, the left column compares a uniform grid representation at four levels (0-3) in the ``Uniform Grid'' column against a multiresolution grid representation at thirteen levels (0-12) in the ``Ytini Method'' column. More than four levels could not easily be represented in a uniform grid due to memory limitations. Thus, the values in the left column are likely to be vastly higher if all thirteen levels could be represented in the uniform grid. A concrete example of this is shown with the number of voxels for levels 0-3 as compared to the number of voxels for levels 0-12. The multiresolution renderings are shown in the bottom half of Figure \ref{fig:renderings}.
\end{table}

\clearpage

\clearpage
\begin{appendix}
\section{A. Python Code} 
Example Python script to read AMR data and write OpenVDB files. Modify the marked lines of code to work with other datasets.
\lstset{language=Python} 
\begin{lstlisting}
import pyopenvdb as vdb
import yt

################### Modify these values to point to your own data ###################
dataFilePath = "/home/kalina/data/raw/enzo_tiny_cosmology/DD0010/DD0010"            #
outFileDir = "/home/kalina/data/vdb/"                                               #
variable = "Density"                                                                #
isFlash = False                                                                     #
#####################################################################################

ds = yt.load(dataFilePath)

minLevel = 0
maxLevel = ds.index.max_level
largestVSize = None

if isFlash:
    ds.periodicity = (True, True, True)

for level in range(minLevel, maxLevel+1):

    gs = ds.index.select_grids(level)

    maskCube = vdb.FloatGrid()
    dataCube = vdb.FloatGrid()

    for index in range(len(gs)):

        subGrid = gs[index]
        subGridVar = subGrid[variable]
        subGridVarGZ = subGrid.retrieve_ghost_zones( n_zones=1, fields=variable)[variable]
        mask = subGrid.child_mask
        ijkout = subGrid.get_global_startindex()

        maskCube.copyFromArray(mask,         ijk=(ijkout[0],ijkout[1],ijkout[2]))
        dataCube.copyFromArray(subGridVarGZ, ijk=(ijkout[0],ijkout[1],ijkout[2]))    

    resolution = ds.domain_dimensions*ds.refine_by**level
    vSize = 1/float(resolution[0])
    
    if level==minLevel:
    	largestVSize = vSize
    
    dataMatrix = [ [vSize,0,0,0], [0,vSize,0,0], [0,0,vSize,0],  \
    		 [-vSize/2-largestVSize, -vSize/2-largestVSize, -vSize/2-largestVSize,1] ]
    maskMatrix = [ [vSize,0,0,0], [0,vSize,0,0], [0,0,vSize,0],  \
    		 [ vSize/2-largestVSize, vSize/2-largestVSize, vSize/2-largestVSize,1] ]
    dataCube.transform = vdb.createLinearTransform(dataMatrix) 
    maskCube.transform = vdb.createLinearTransform(maskMatrix)

    output = []
    dataCube.name = "density"
    maskCube.name = "mask"
    output.append(maskCube)
    output.append(dataCube)
    outFileName = "%s/%s_level%d.vdb" % (outFileDir, variable, level)
    vdb.write(outFileName, grids=output)


\end{lstlisting}

\section{B. Detailed Houdini Workflow} 

Here we present a step-by-step workflow importing a sample AMR dataset into \textit{Houdini}. The implementation of this workflow takes two parts - external data processing in Python, and setting up the shader in \textit{Houdini}. The detailed reasoning behind these steps is described above in Section \ref{section:amrHoudini}. The description here builds on the assumption of basic knowledge of the \textit{Houdini} user interface, as described in our previous paper \citep{naiman2017}.

Begin by installing \textit{Houdini}, \textit{yt}, and pyopenvdb as described in the ``Install pyopenvdb'' tutorial at \textit{www.ytini.com}\footnote{http://ytini.com/tutorials/tutorial\_vdbInstall.html}. Copy the Python code provided in the Appendix, and modify the marked variables, \lstinline[basicstyle=\ttfamily]|dataFilePath|, \lstinline[basicstyle=\ttfamily]|outFileDir|, \lstinline[basicstyle=\ttfamily]|variable|, and \lstinline[basicstyle=\ttfamily]|isFlash|, to match your FLASH, Enzo, or Athena dataset, or download the sample data provided at \textit{www.ytini.com}\footnote{http://ytini.com/downloaddata.html}. The Python code is also available for download at \textit{www.ytini.com} in the ``AMR Data'' tutorial, with line-by-line comments and explanations. Running this Python script will create OpenVDB .vdb files in the directory defined by \lstinline[basicstyle=\ttfamily]|outFileDir|.

\begin{figure*}
\centering
\includegraphics[width=1.0\textwidth]{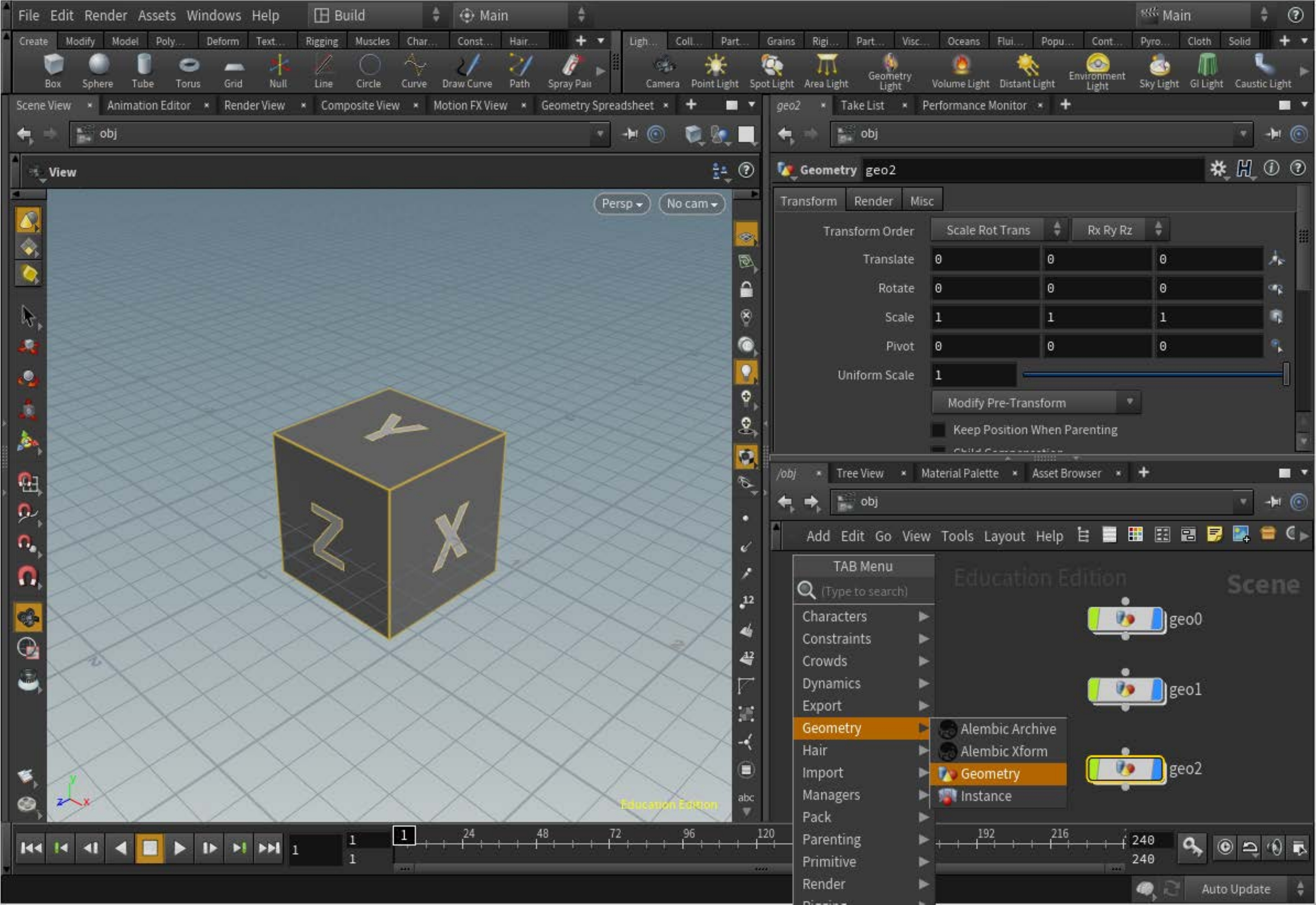}
\caption{\textit{Houdini}'s TAB Menu is used to create a Geometry node for each level of refinement of the AMR dataset.}
\label{fig:tutorial1}
\end{figure*}

\begin{figure*}
\centering
\includegraphics[width=1.0\textwidth]{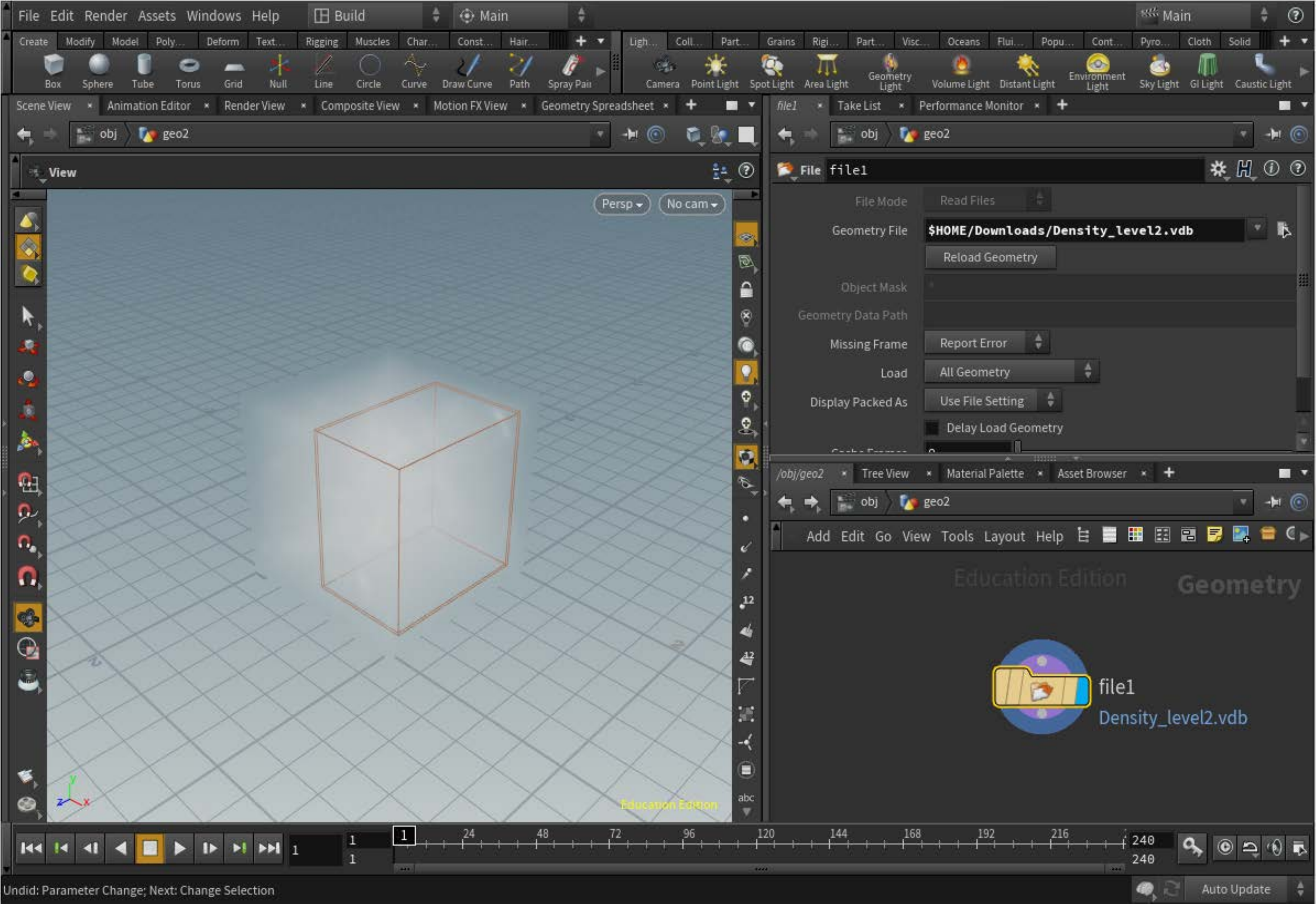}
\caption{One can double-click to step inside each Geometry node created in Figure \ref{fig:tutorial1} and then modify the path in the File node to point to the .vdb file corresponding to each level.}
\label{fig:tutorial2}
\end{figure*}

Open \textit{Houdini}. To import the data, use the TAB Menu to create a Geometry node for each level, as shown in Figure \ref{fig:tutorial1} (see the "Render an Image" tutorial on \textit{www.ytini.com}\footnote{http://ytini.com/tutorials/tutorial\_renderAnImageH16.html} for more details on this process). Step inside the Geometry node and modify the path in the File node to point to the level 0 .vdb file created by running the Python script. Create Geometry nodes for each level, pointing each internal File node to its respective level's .vdb file, as shown in Figure \ref{fig:tutorial2}.

\begin{figure*}
\centering
\includegraphics[width=1.0\textwidth]{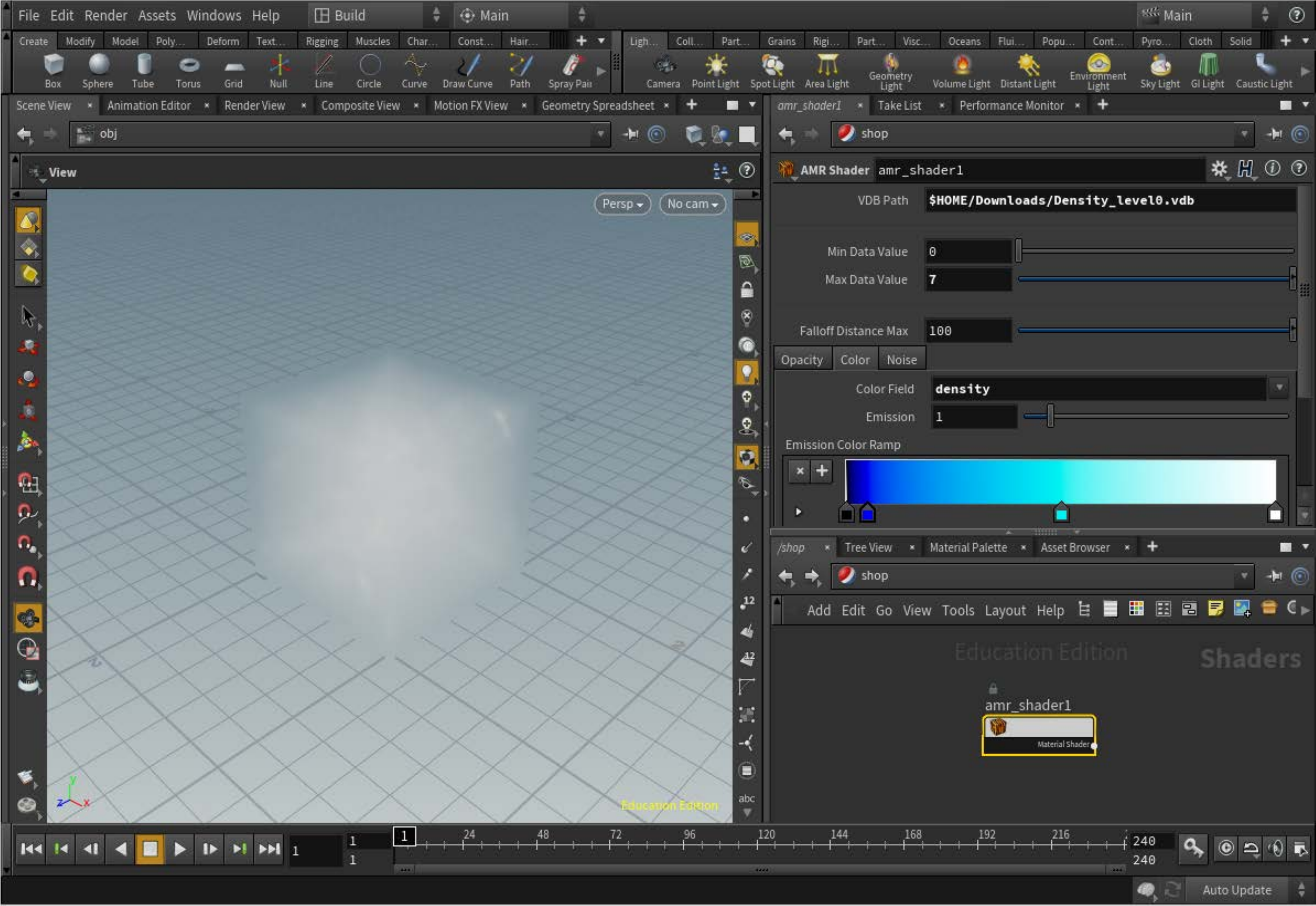}
\caption{ The shader downloaded from\textit{ www.ytini.com} -\textit{ http://ytini.com/docs-assets/tutorial\_amr/amr\_shader.hdanc} - creates a color slider and various tunable parameters to change the rendered appearance of our dataset.  To use once imported, set the VDB Path parameter to point to the .vdb file location.  Then set the Min and Max Data Value fields to the minimum and maximum data values. In the Color tab, set Color Field to ``density''. Create a color map in the Emission Color Ramp bar.}
\label{fig:tutorial3}
\end{figure*}

Download the AMR shader digital asset from \textit{www.ytini.com}\footnote{http://ytini.com/docs-assets/tutorial\_amr/amr\_shader.hdanc}. The internal workings of this custom shader are described above in Section \ref{section:amrHoudini} (see section 3 of \cite{naiman2017} for more details about shaders in \textit{Houdini} in general). To load the shader into the \textit{Houdini} scene, navigate to \lstinline[basicstyle=\ttfamily]|File -> Import -> Houdini Digital Asset|. Select the downloaded \lstinline[basicstyle=\ttfamily]|amr_shader.hdanc| file and click ``Install and Create''. This will create a node called \lstinline[basicstyle=\ttfamily]|amr_shader1| in the \lstinline[basicstyle=\ttfamily]|/shop/| network. Set the shader parameter ``VDB Path'' to point to the level 0 volume. Set ``Min Data Value'' and ``Max Data Value'' to the minimum and maximum data values, respectively. In the Color tab, set ``Color Field'' to ``density''. To optionally color the volume, create a color map in the ``Emission Color Ramp'' bar, which will result in a screen similar to the one shown in Figure \ref{fig:tutorial3}. 

\begin{figure*}
\centering
\includegraphics[width=1.0\textwidth]{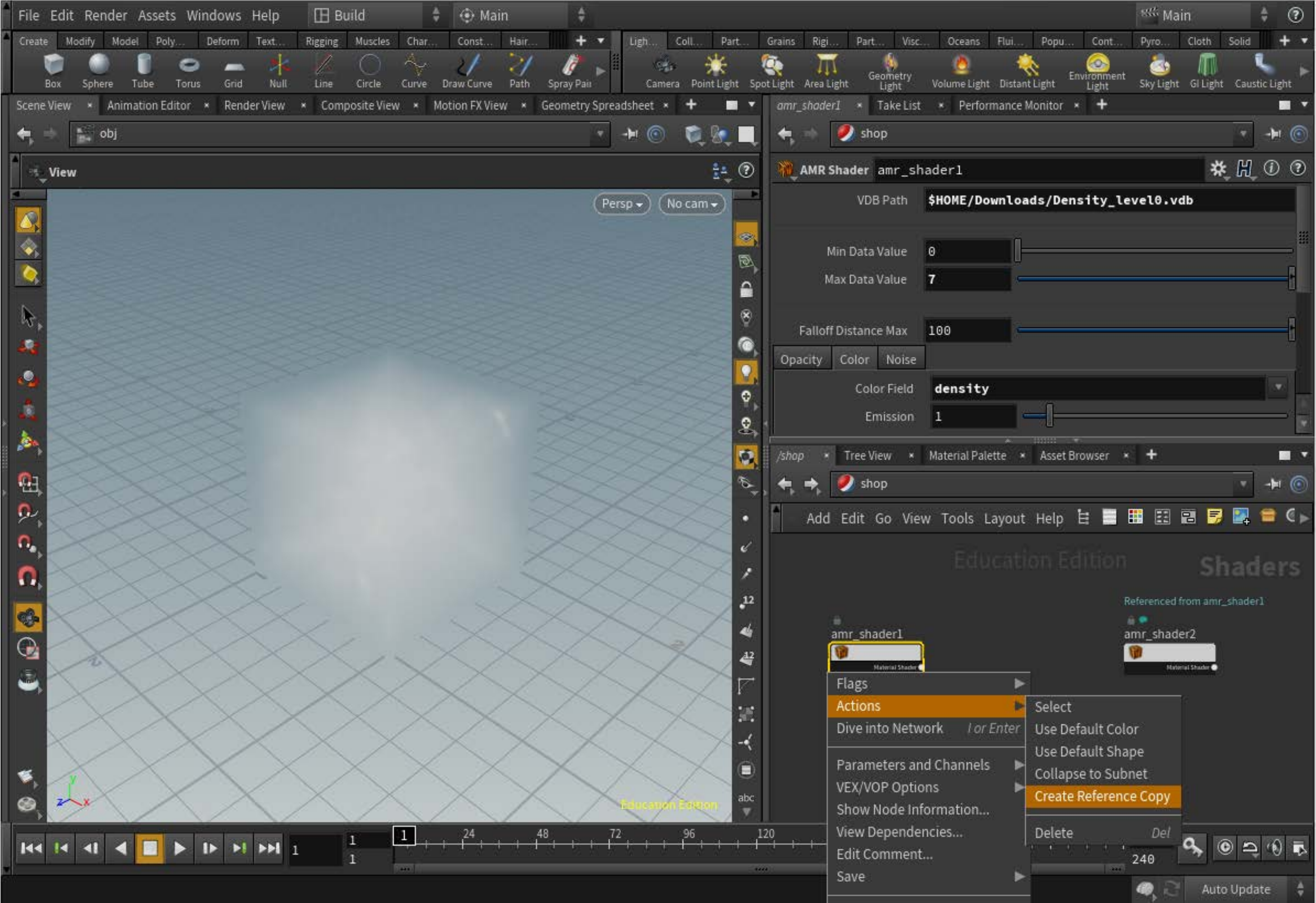}
\caption{To apply the downloaded shader to each level of refinement, create a reference copy of the first shader.}
\label{fig:tutorial4}
\end{figure*}

\begin{figure*}
\centering
\includegraphics[width=1.0\textwidth]{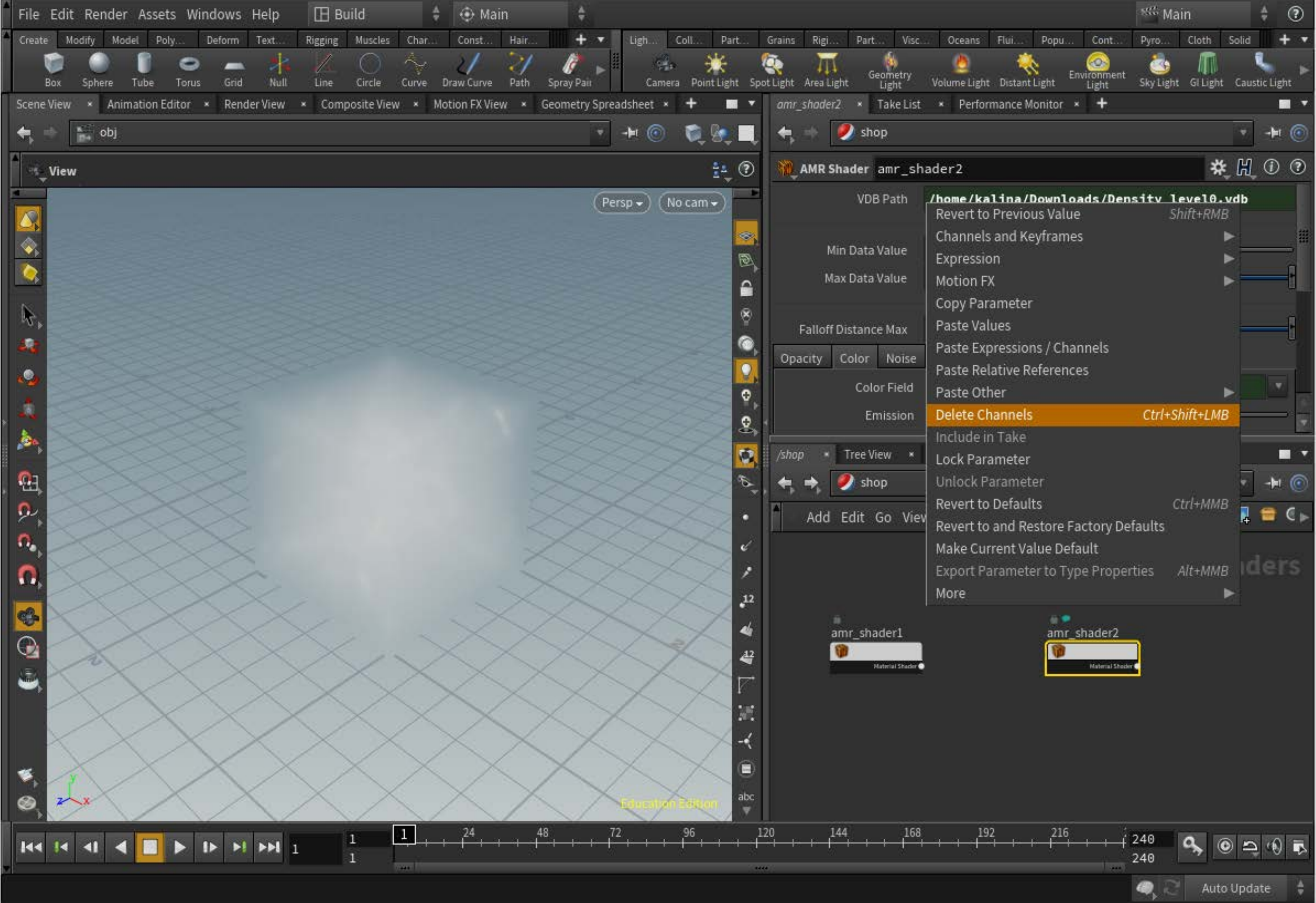}
\caption{In order to apply each reference shader to the appropriate refinement level volume, navigate to each individual shader and delete the automatically-populated VDB Path channel and change the text to point to the appropriate level .vdb file.}
\label{fig:tutorial5}
\end{figure*}

Next, create separate shaders for each AMR level volume. Each of these shaders must have identical values, only differing in file name under ``VDB Path''. To avoid having to maintain multiple copies of similar shaders, create a reference copy of the level 0 shader to link parameter values between all copies. To make a reference copy of the first shader, right-click on the \lstinline[basicstyle=\ttfamily]|amr_shader1| node and select \lstinline[basicstyle=\ttfamily]|Actions -> Create Reference Copy|, as shown in Figure \ref{fig:tutorial4}. This creates a node \lstinline[basicstyle=\ttfamily]|amr_shader2|, which links each parameter to those in \lstinline[basicstyle=\ttfamily]|amr_shader1|. In this new copy node, right click on the ``VDB Path'' field and select ``Delete Channels'' as per Figure \ref{fig:tutorial5}. This will unlink this field from \lstinline[basicstyle=\ttfamily]|amr_shader1|, and will allow the text to be editable. Update the path to point to the level 1 volume. Repeat these steps to make referenced copies of the shader for each AMR level.

\begin{figure*}
\centering
\includegraphics[width=1.0\textwidth]{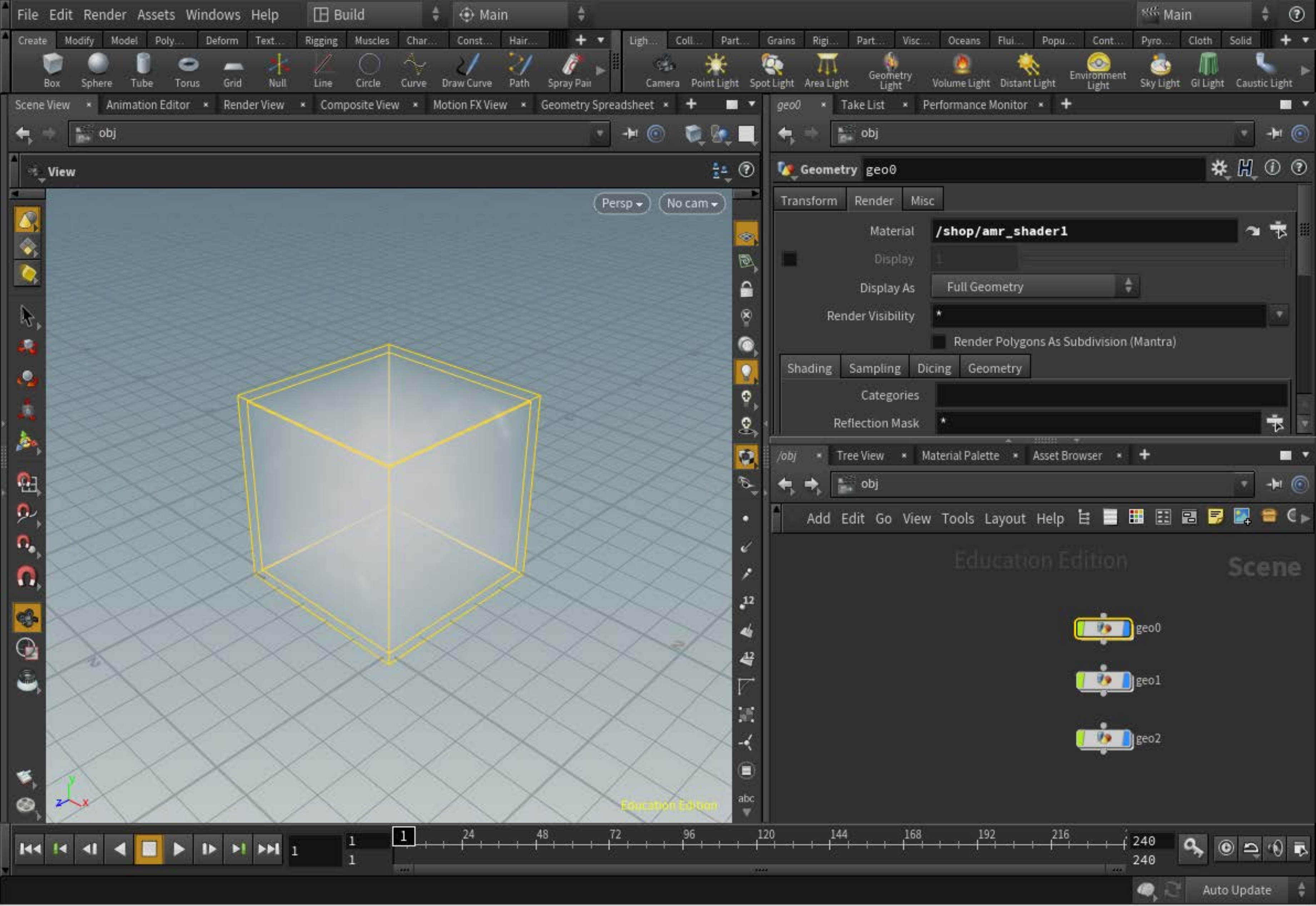}
\caption{In order to link each OpenVDB object representing each level with its appropriate shader, return to the \lstinline[basicstyle=\ttfamily]|/obj/| network view. For each Geometry node, navigate to the Render parameter tab and set the Material to the respective level's shader.}
\label{fig:tutorial6}
\end{figure*}

Return to the \lstinline[basicstyle=\ttfamily]|/obj/| network. For each Geometry node, navigate to the ``Render'' parameters tab, and set the ``Material'' to the respective level's \lstinline[basicstyle=\ttfamily]|amr_shader|. Your screen should look like Figure \ref{fig:tutorial6}.

\begin{figure*}
\centering
\includegraphics[width=1.0\textwidth]{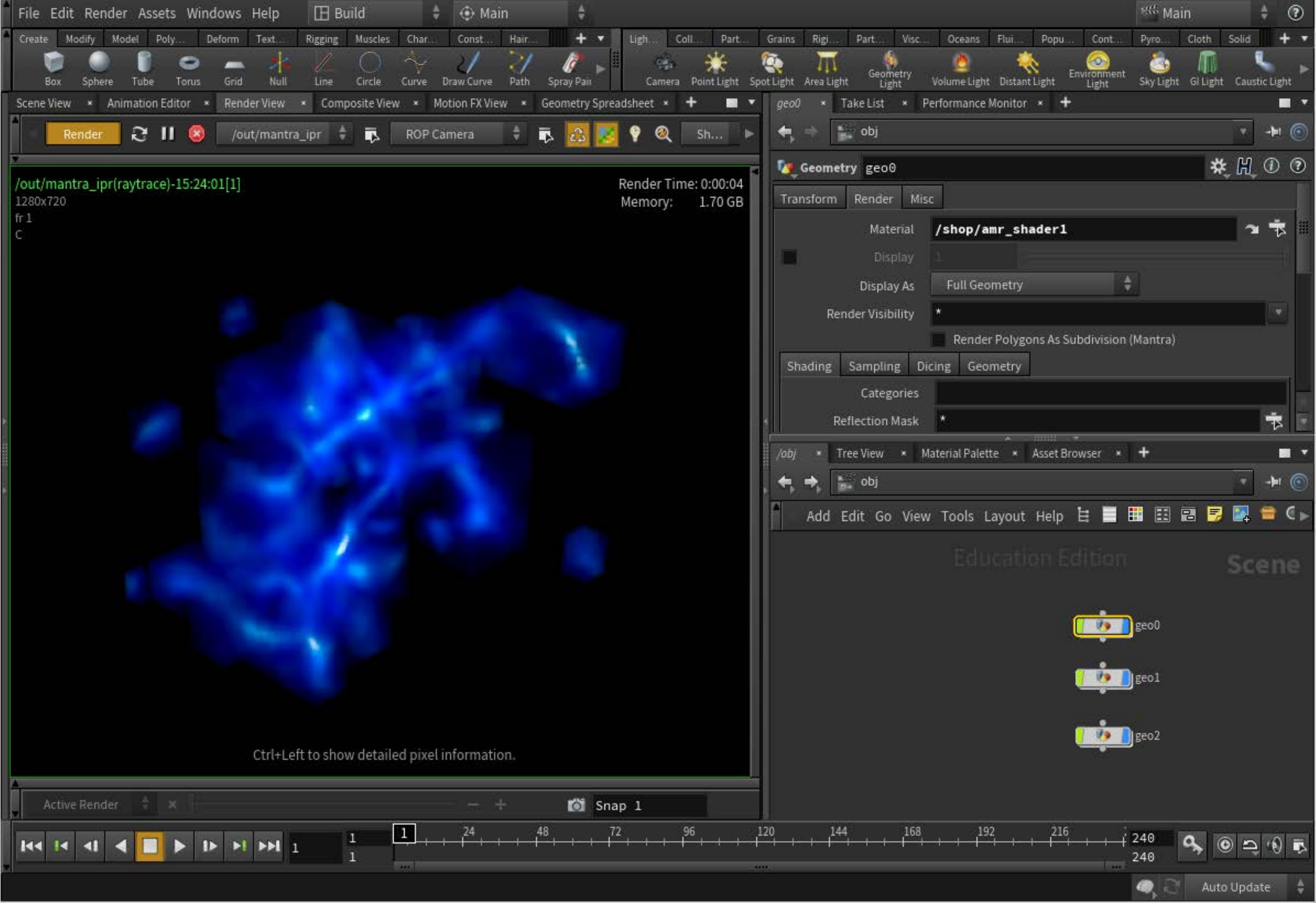}
\caption{Click on the Render button in the Render View tab to view the rendering.}
\label{fig:tutorial7}
\end{figure*}

In the Viewport, click on the ``Render View'' tab and click on the ``Render'' button to create an image. If needed, return to the shader to modify the color map, and check back in the Render View tab to see the changes. The final image will look similar to the one shown in Figure \ref{fig:tutorial7}. Right-click on the image and select ``Save Frame'' in order to save the final result. A more detailed tutorial with a full explanation can be found at\textit{ www.ytini.com}\footnote{http://ytini.com/tutorials/tutorial\_amr.html}.

\end{appendix}

\clearpage

\clearpage
\bibliographystyle{apj}

\bibliography{bib_amr}

\begin{thebibliography}{}
\expandafter\ifx\csname natexlab\endcsname\relax\def\natexlab#1{#1}\fi

\bibitem[{{Ahrens} {et~al.}(2005){Ahrens}, {Geveci}, \& {Law}}]{paraview}
{Ahrens}, J., {Geveci}, B., \& {Law}, C. 2005, {Visualization Handbook}, 717

\bibitem[{{Arroio}(2010)}]{arroio2010}
{Arroio}, A. 2010, Science Education International, 21, 131

\bibitem[{{Barnes} \& {Fluke}(2008)}]{barnes2008}
{Barnes}, D.~G., \& {Fluke}, C.~J. 2008, New Astronomy, 13, 599

\bibitem[{Berger \& Colella(1989)}]{amr}
Berger, M., \& Colella, P. 1989, Journal of Computational Physics, 82, 64

\bibitem[{Borkiewicz {et~al.}(2017)Borkiewicz, Christensen, \&
  Stone}]{borkiewicz2017}
Borkiewicz, K., Christensen, A.~J., \& Stone, J.~E. 2017, in SIGGRAPH Asia 2017
  Courses, SA '17 (New York, NY, USA: ACM), 3:1--3:122

\bibitem[{Borland \& Ii(2007)}]{rainbowmap1}
Borland, D., \& Ii, R. M.~T. 2007, IEEE Computer Graphics and Applications, 27,
  14

\bibitem[{{Bryan} {et~al.}(2014){Bryan}, {Norman}, {O'Shea}, {Abel}, {Wise},
  {Turk}, {Reynolds}, {Collins}, {Wang}, {Skillman}, {Smith}, {Harkness},
  {Bordner}, {Kim}, {Kuhlen}, {Xu}, {Goldbaum}, {Hummels}, {Kritsuk}, {Tasker},
  {Skory}, {Simpson}, {Hahn}, {Oishi}, {So}, {Zhao}, {Cen}, {Li}, \& {Enzo
  Collaboration}}]{enzo2013}
{Bryan}, G.~L., {Norman}, M.~L., {O'Shea}, B.~W., {et~al.} 2014, The
  Astrophysical Journal Supplement Series, 211, 19

\bibitem[{Cawthon \& Moere(2007)}]{cawthon}
Cawthon, N., \& Moere, A.~V. 2007, in 2007 11th International Conference
  Information Visualization (IV '07), 637--648

\bibitem[{Chen(2005)}]{chen}
Chen, C. 2005, IEEE Computer Graphics and Applications, 25, 12

\bibitem[{Childs {et~al.}(2012)Childs, Brugger, Whitlock, Meredith, Ahern,
  Pugmire, Biagas, Miller, Harrison, Weber, Krishnan, Fogal, Sanderson, Garth,
  Bethel, Camp, R\"{u}bel, Durant, Favre, \& Navr\'{a}til}]{visit}
Childs, H., Brugger, E., Whitlock, B., {et~al.} 2012, in {High Performance
  Visualization--Enabling Extreme-Scale Scientific Insight}, 357--372

\bibitem[{{Dubcek} {et~al.}(2003){Dubcek}, {Moshier}, \& {Boss}}]{dubcek2003}
{Dubcek}, L., {Moshier}, S., \& {Boss}, J. 2003, {Fantastic Voyagers: Learning
  Science through Science Fiction}

\bibitem[{Eddins(2014)}]{rainbowmap2}
Eddins, S. 2014, MathWorks Technical Articles and Newsletters, 25

\bibitem[{{Fryxell} {et~al.}(2000){Fryxell}, {Olson}, {Ricker}, {Timmes},
  {Zingale}, {Lamb}, {MacNeice}, {Rosner}, {Turuan}, \& {Tufo}}]{flash2000}
{Fryxell}, B., {Olson}, K., {Ricker}, P., {et~al.} 2000, The Astrophysical
  Journal Supplement Series, 131, 273

\bibitem[{{Goodman}(2012)}]{goodman2012}
{Goodman}, A.~A. 2012, Astronomische Nachrichten, 333, 505

\bibitem[{James {et~al.}(2015)James, von Tunzelmann, Franklin, \&
  Thorne}]{interstellar}
James, O., von Tunzelmann, E., Franklin, P., \& Thorne, K.~S. 2015, Classical
  and Quantum Gravity, 32, 065001

\bibitem[{Kaehler \& Abel(2013)}]{bespoke2}
Kaehler, R., \& Abel, T. 2013, in Visualization and Data Analysis 2013, ed.
  P.~C. Wong, D.~L. Kao, M.~C. Hao, C.~Chen, \& C.~G. Healey ({SPIE})

\bibitem[{Kahler {et~al.}(2003)Kahler, Simon, \& Hege}]{bespoke1}
Kahler, R., Simon, M., \& Hege, H.~. 2003, IEEE Transactions on Visualization
  and Computer Graphics, 9, 341

\bibitem[{{Kent}(2015)}]{kent2015}
{Kent}, B.~R. 2015, {3D Scientific Visualization with Blender} (Morgan \&
  Claypool)

\bibitem[{Kim \& Park(2013)}]{important1}
Kim, H., \& Park, J.~W. 2013, in SIGGRAPH Asia 2013 Art Gallery, SA '13 (New
  York, NY, USA: ACM), 20:1--20:7

\bibitem[{Laramee {et~al.}(2014)Laramee, Carr, Chen, Hauser, Linsen, Mueller,
  Natarajan, Obermaier, Peikert, \& Zhang}]{laramee2014}
Laramee, R.~S., Carr, H., Chen, M., {et~al.} 2014, Future Challenges and
  Unsolved Problems in Multi-field Visualization, ed. C.~D. Hansen, M.~Chen,
  C.~R. Johnson, A.~E. Kaufman, \& H.~Hagen (London: Springer London), 205--211

\bibitem[{Li(2018)}]{visart}
Li, Q. 2018, Visual Communication, 17, 299

\bibitem[{Mellinger(2009)}]{allsky}
Mellinger, A. 2009, Publications of the Astronomical Society of the Pacific,
  121, 1180

\bibitem[{Moere \& Purchase(2011)}]{important2}
Moere, A.~V., \& Purchase, H. 2011, Information Visualization, 10, 356

\bibitem[{Moreland(2016)}]{rainbowmap3}
Moreland, K. 2016, Electronic Imaging, 2016

\bibitem[{Museth {et~al.}(2013)Museth, Lait, Johanson, Budsberg, Henderson,
  Alden, Cucka, Hill, \& Pearce}]{museth2013}
Museth, K., Lait, J., Johanson, J., {et~al.} 2013, in ACM SIGGRAPH 2013
  Courses, SIGGRAPH '13 (New York, NY, USA: ACM), 19:1--19:1

\bibitem[{{Naiman}(2016)}]{naiman2016}
{Naiman}, J.~P. 2016, Astronomy and Computing, 15, 50

\bibitem[{{Naiman} {et~al.}(2017){Naiman}, {Borkiewicz}, \&
  {Christensen}}]{naiman2017}
{Naiman}, J.~P., {Borkiewicz}, K., \& {Christensen}, A.~J. 2017, Publications
  of the Astronomical Society of the Pacific, 129, 058008

\bibitem[{Nelson {et~al.}(2011)Nelson, Brown, Brun, Miesch, \& Toomre}]{miesch}
Nelson, N.~J., Brown, B.~P., Brun, A.~S., Miesch, M.~S., \& Toomre, J. 2011,
  The Astrophysical Journal Letters, 739, L38

\bibitem[{{O'Shea} {et~al.}(2015){O'Shea}, {Wise}, {Xu}, \&
  {Norman}}]{oshea2015}
{O'Shea}, B.~W., {Wise}, J.~H., {Xu}, H., \& {Norman}, M.~L. 2015, The
  Astrophysical Journal Letters, 807, L12

\bibitem[{Pandey {et~al.}(2014)Pandey, Manivannan, Nov, Satterthwaite, \&
  Bertini}]{pandey}
Pandey, A.~V., Manivannan, A., Nov, O., Satterthwaite, M., \& Bertini, E. 2014,
  IEEE Transactions on Visualization and Computer Graphics, 20, 2211

\bibitem[{Pesnell {et~al.}(2012)Pesnell, Thompson, \& Chamberlin}]{sdo}
Pesnell, W.~D., Thompson, B.~J., \& Chamberlin, P.~C. 2012, Solar Physics, 275,
  3

\bibitem[{{Punzo} {et~al.}(2015){Punzo}, {van der Hulst}, {Roerdink},
  {Oosterloo}, {Ramatsoku}, \& {Verheijen}}]{punzo2015}
{Punzo}, D., {van der Hulst}, J.~M., {Roerdink}, J.~B.~T.~M., {et~al.} 2015,
  Astronomy and Computing, 12, 86

\bibitem[{Rempel \& Cheung(2014)}]{rempel}
Rempel, M., \& Cheung, M. C.~M. 2014, The Astrophysical Journal, 785, 90

\bibitem[{{Serra} \& {Arroio}(2008)}]{serra2008}
{Serra}, G.~M.~D., \& {Arroio}, A. 2008, XII IOSTE Symposium Proceedings: The
  use of Science and Technology Education for Peace and Sustainable
  Development, 1185

\bibitem[{{Stone} {et~al.}(2008){Stone}, {Gardiner}, {Teuben}, {Hawley}, \&
  {Simon}}]{athena2008}
{Stone}, J.~M., {Gardiner}, T.~A., {Teuben}, P., {Hawley}, J.~F., \& {Simon},
  J.~B. 2008, The Astrophysical Journal Supplement Series, 178, 137

\bibitem[{Taylor(2015)}]{taylor2015}
Taylor, R. 2015, Astronomy and Computing, 13, 67

\bibitem[{{Turk} {et~al.}(2011){Turk}, {Smith}, {Oishi}, {Skory}, {Skillman},
  {Abel}, \& {Norman}}]{turk2011}
{Turk}, M.~J., {Smith}, B.~D., {Oishi}, J.~S., {et~al.} 2011, The Astrophysical
  Journal Supplement Series, 192, 9

\bibitem[{Vincenti {et~al.}(1971)Vincenti, , \& Traugott}]{raytracingastro}
Vincenti, W.~G., , \& Traugott, S.~C. 1971, Annual Review of Fluid Mechanics,
  3, 89

\bibitem[{{Vogt} {et~al.}(2016){Vogt}, {Owen}, {Verdes-Montenegro}, \&
  {Borthakur}}]{vogt2016}
{Vogt}, F.~P.~A., {Owen}, C.~I., {Verdes-Montenegro}, L., \& {Borthakur}, S.
  2016, The Astrophysical Journal, 818, 115

\bibitem[{Wald {et~al.}(2017)Wald, Brownlee, Usher, \& Knoll}]{cpuamr}
Wald, I., Brownlee, C., Usher, W., \& Knoll, A. 2017, in SIGGRAPH Asia 2017
  Symposium on Visualization, SA '17 (New York, NY, USA: ACM), 9:1--9:8

\bibitem[{Welbourne \& Grant(2016)}]{welbourne}
Welbourne, D.~J., \& Grant, W.~J. 2016, Public Understanding of Science, 25,
  706

\bibitem[{{Wise} {et~al.}(2012){Wise}, {Turk}, {Norman}, \& {Abel}}]{wise}
{Wise}, J.~H., {Turk}, M.~J., {Norman}, M.~L., \& {Abel}, T. 2012, The
  Astrophysical Journal, 745, 50

\end{thebibliography}

\end{document}